\documentclass[12pt]{article}

\usepackage[margin=1in]{geometry}

\usepackage{amsfonts, amssymb, amsmath, amsthm, multicol}
\usepackage[colorlinks=true, pdfstartview=FitV, linkcolor=blue,
            citecolor=blue, urlcolor=blue, bookmarksdepth=3]{hyperref}
\usepackage{amsthm,amsmath,amssymb,amsfonts}
\usepackage{setspace,graphicx}
\usepackage{url} 
\usepackage{enumitem}
\usepackage[capitalize,nameinlink,noabbrev]{cleveref}
\usepackage[table,dvipsnames]{xcolor}

\numberwithin{equation}{section}
\theoremstyle{plain}
\newtheorem{theorem}{Theorem}[section]

\usepackage{algorithm}
\usepackage[noend]{algpseudocode}		

\newcommand{\RR}{\mathbb{R}}

\newcommand{\cC}{\mathcal{C}}
\newcommand{\cN}{\mathcal{N}}
\newcommand{\spq}{X}
\newcommand{\spv}{Y}
\newcommand{\bq}{\mathbf{q}}
\newcommand{\bv}{\mathbf{v}}
\newcommand{\bz}{\mathbf{z}}

\newcommand{\btq}{\tilde{\mathbf{q}}}
\newcommand{\Vker}{\mathcal{V}}
\newcommand{\Pker}{P}
\newcommand{\cM}{\mathcal{M}}

\newcommand{\arFn}{\hat{\alpha}}
\newcommand{\dimx}{N}

\newcommand{\dmu}{\pi}
\newcommand{\dnu}{h}
\newcommand{\dVker}{r}

\newcommand{\Ham}{H} 
\newcommand{\dt}{\delta} 

\newcommand{\Pot}{\Phi}

\newcommand{\nsn}{\eta}
\newcommand{\datan}{y}
\newcommand{\ns}{\boldsymbol{\nsn}}
\newcommand{\data}{\mathbf{\datan}}
\newcommand{\fwd}{\mathcal{G}}
\newcommand{\mpr}{\mu_0}
\newcommand{\mns}{\rho_{\ns}}
\newcommand{\mps}{\mu_{\data}}
\newcommand{\scf}{\theta}
\newcommand{\vf}{\mathbf{v}}
\newcommand{\x}{\mathbf{x}}
\newcommand{\diff}{\kappa}

\newcommand{\unk}{\vf}

\newcommand{\agp}{f}
\newcommand{\da}{\delta_a}
\newcommand{\db}{\delta_b}

 \usepackage{authblk}
 
 \title{Sacred and Profane:
  from the Involutive Theory
  of MCMC to Helpful Hamiltonian Hacks\footnote{To appear in the Handbook of Markov chain Monte Carlo, 2nd Edition.}}
\author[1]{Nathan E. Glatt-Holtz\footnote{negh@iu.edu}}
\author[2]{Andrew J. Holbrook\footnote{aholbroo@g.ucla.edu}}
\author[3,4]{Justin A. Krometis\footnote{jkrometis@vt.edu}} 
\author[5]{Cecilia F. Mondaini\footnote{cf823@drexel.edu}}
\author[2]{Ami Sheth\footnote{amisheth26@g.ucla.edu}}

\affil[1]{Department of Statistics, Indiana University Bloomington}
\affil[2]{Department of Biostatistics, University of California, Los Angeles}
\affil[3]{National Security Institute, Virginia Tech}
\affil[4]{Department of Mathematics, Virginia Tech}
\affil[5]{Department of Mathematics, Drexel University}

\date{}

\begin{document}

\maketitle

\section{Introduction}

In the first edition of this Handbook, two remarkable chapters
consider seemingly distinct yet deeply connected subjects.  
On the one hand, Geyer's influential and insightful \emph{Introduction to Markov Chain
  Monte Carlo} broadens the scope of typical MCMC introductions to
include a number of ideas from \cite{green1995reversible},
asserting
\begin{quote}
  ``\cite{green1995reversible} generalized the Metropolis-Hastings
  algorithm as much as it can be generalized. Although this
  terminology is not widely used, we say that simulations following
  his scheme use the Metropolis-Hastings-Green algorithm.''
  \cite{geyer2011introduction}
\end{quote}
In its final sections, Geyer's chapter presents an important
generalization of the classical Metropolis-Hastings method, the
Metropolis-Hastings-Green with Jacobians (MHGJ) algorithm.  The MHGJ
framework adapts the Metropolis-Hastings acceptance criterion to
account for proposals involving deterministic mappings $S$ on an
augmented state space.  Crucially, Geyer requires that $S$ be its own
inverse, i.e., an involution, although he does not use this specific
term.

On the other hand, Neal's landmark \emph{MCMC Using Hamiltonian
  Dynamics} presents a comprehensive but intuitive introduction to the
Hamiltonian (hybrid) Monte Carlo (HMC) method of
\cite{duane1987hybrid}.  At the heart of HMC is the integration of an
appropriately chosen Hamiltonian dynamic that determines the proposal
at each step of the chain.  A perfect resolution of this dynamic
exactly conserves the associated energy functional and obviates the
need for a Metropolis correction step.  Since perfect resolutions are
rarely available in practice, one typically resorts to numerical
approximation, and Neal identifies path-reversibility and volume
preservation as two crucial properties a numerical integrator must
retain.  In particular, volume-preservation means that the absolute
value of the determinant of the Jacobian of the numerical map is
unity. Per Neal,
\begin{quote}
  ``If we proposed new states using some arbitrary, non-Hamiltonian,
  dynamics, we would need to compute the determinant of the Jacobian
  matrix for the mapping the dynamics defines, which might well be
  infeasible.'' \cite{neal2011mcmc}
\end{quote}
Just like Geyer, Neal understands that introducing deterministic
dynamics within the MCMC proposal step requires adapting the original
Metropolis-Hastings acceptance criterion via Jacobian corrections.  In
his path-reversibility requirement, Neal also hints at an underlying
involutive structure that HMC must maintain.  Notwithstanding this
common thread, Neal stops short of placing HMC within an overarching
MHGJ-type framework.

Recently, however, a collection of parallel works
\cite{andrieu2020general, glatt2023accept, neklyudov2020involutive,
  glatt2024parallel} make the connection between MHGJ and HMC fully
explicit.  Moreover, these newer works extend the measure-theoretic
foundations advocated by \cite{green1995reversible} and \cite{
  Tierney1998} to encompass a more general, involutive, extended phase
space formalism.  Counter to the assertion that one cannot further
generalize the Metropolis-Hastings algorithm beyond
\cite{green1995reversible}, these recent works reveal opportunities to
push the MCMC paradigm forward, both by establishing new theoretical
results and by providing a framework for novel algorithms. Among other
contributions, \cite{glatt2023accept} and \cite{glatt2024parallel}
\begin{enumerate}[label={(\arabic{enumi})}]
\item provide a complete and unified treatment of surrogate-trajectory
  methods that sidestep expensive or intractable gradient computations
  that may occur within HMC;
\item extend these methods to the setting of Hilbert space MCMC
  algorithms that partially beat the curse of dimensionality for
  certain infinite-dimensional target measures;
\item develop multiproposal MCMC algorithms that generate entire
  clouds of proposals at each step in order to improve sampling for
  targets with difficult geometries.
\end{enumerate}
Notably, Hilbert methods (2) have been crucial in the development of
the Bayesian approach to PDE inverse problems
\cite{stuart2010inverse, dashti2017bayesian, borggaard2020bayesian},
and the multiproposal approach (3) is amenable to novel within-chain
parallelization approaches facilitated by contemporary computational
hardware.

In the following, we review an extended involutive MCMC framework that
includes single and multiple proposal algorithms on general state
spaces and relate this framework to Metropolis-Hastings, MHGJ and
HMC-type algorithms. We highlight this framework's applicability to
surrogate-trajectory HMC, algorithms for Bayesian inference on Hilbert
spaces and a recent, simplified, multiproposal MCMC algorithm.  Next,
we focus on surrogate-trajectory HMC within three short case studies
involving Bayesian inference.  An application to phylogenetic
continuous-time Markov chain models demonstrates the promise of
surrogate-trajectory methods, while an application to Bayesian
multidimensional scaling provides a negative example. A third
application to the inversion of nonlinear partial differential
equations demonstrates the applicability of surrogate-trajectory HMC
(and a new multiproposal MCMC algorithm) to target distributions
defined on Hilbert spaces.  We conclude with avenues for future
research.

\section{The Involutive Framework and its Extensions}
\label{sec:invol}

\begin{algorithm}[!t]
	\caption{\cite{glatt2024parallel}}
	\begin{algorithmic}[1]
          \State Select:
          \begin{itemize}
          \item[(i)] An auxiliary measurable space $(Y, \Sigma_Y)$.
          \item[(ii)] A proposal kernel $\Vker: X\times \Sigma_Y \to [0,1]$.
          \item[(iii)] A proposal count $p \geq 1$.
          \item[(iv)] Measurable mappings $S_j: X \times Y \to X \times Y$ for $j = 0,\ldots, p$.
          \item[(v)] Acceptance probabilities $\arFn_j:X \times Y \to [0,1]$ with $\sum_{j =0}^p \arFn_j \equiv 1$.
          \end{itemize}  
          \State Start from an initial point $\bq^{(0)} \in \spq$.
          \For{$k \geq 0$}
          \State Sample $\bv^{(k+1)} \sim \Vker(\bq^{(k)}, \cdot)$.
          \State Compute $S_j(\bq^{(k)}, \bv^{(k+1)})$, for $j = 0,1,\ldots,p$.
          \State Set $\bq^{(k+1)}$ among $(\Pi_1 
          S_0(\bq^{(k)}, \bv^{(k+1)}), \ldots, \Pi_1 S_p(\bq^{(k)}, \bv^{(k+1)}))$,
          where $\Pi_1(\bq, \bv) = \bq$,
                    \par
          with probabilities
          $(\arFn_0(\bq^{(k)}, \bv^{(k+1)}), \ldots,  \arFn_p(\bq^{(k)}, \bv^{(k+1)}))$.
          \EndFor
	\end{algorithmic}\label{alg:main:master}
\end{algorithm}

The recent works \cite{andrieu2020general}, \cite{glatt2023accept},
\cite{neklyudov2020involutive} and \cite{glatt2024parallel} provide
a master algorithm under which essentially all of the known
(reversible) Metropolis-Hastings-type sampling methods fall out as
special cases.  Here, we demonstrate how the reversibility result,
\cref{thm:mprop} of \cite{glatt2024parallel}, provides a basis for
understanding many well-known methods while allowing the
identification of new and useful algorithms.  Throughout this section
we suppose that $\mu$ is a probability measure on a measurable state
space $(\spq,\Sigma_\spq)$. With the aim of resolving $\mu$, we
consider \cref{alg:main:master}, where each iterative step starting
from a given state $\bq \in \spq$ has an associated transition kernel
given by
\begin{align}\label{eq:main:master:ker}
    \Pker(\bq,d\btq) 
    = \sum_{j = 0}^p \int_\spv \arFn_j(\bq,\bv) \delta_{\Pi_1  S_j(\bq,\bv)}(d\btq)
    \Vker(\bq,d \bv).
\end{align}
Of course, without any constraints on $(\Vker, S, \arFn)$, we have no
reason to believe that a generic $P$ will hold $\mu$
invariant.\footnote{Indeed, note that any Markov kernel $\Pker$ on a
  measurable space $\spq$ can be written as in
  \eqref{eq:main:master:ker} by choosing $\spv = \spq$, $p = 1$,
  $\arFn_0 \equiv 1$ (so that $\arFn_1 \equiv 0$),
  $S_0(\bq, \bv) = (\bv, \bq)$, $S_1$ as any mapping, and
  $\Vker = \Pker$.  }

As a simple preliminary illustration of how \eqref{eq:main:master:ker}
includes well-known algorithms, we obtain the original Random Walk
Metropolis (RWM) method as follows. Consider a target measure
continuously distributed on $\RR^N$, $\mu(d \bq) = \dmu (\bq) d\bq$,
where we employ a symmetric kernel $\dVker(\bq, \btq)d\btq$ for our
proposal step.  In this context, the RWM kernel arises by taking $X = Y= \RR^N$, with the standard Borel
$\sigma$-algebra, $p=1$, $\Vker(\bq, d\btq) = \dVker(\bq, \btq)d\btq$,
$S_0(\bq, \bv) = (\bq,\bv)$, $S_1(\bq,\bv) = (\bv,\bq)$ and finally
$\arFn_1(\bq,\bv) = 1 \wedge(\pi(\bv)/\pi(\bq))$ so that
$\arFn_0(\bq,\bv) =1 - \arFn_1(\bq,\bv)$. In \cref{sec:MH:classic}, \cref{sec:Ham:Methods},
\cref{eq:sj:split}, \cref{sec:Hilbert} and \cref{sec:Multiprop}, we demonstrate that all of the classical
Metropolis-Hastings, HMC, Hilbert space methods, multiproposal methods
and numerous other possibilities all fall out of
\cref{alg:main:master} and \eqref{eq:main:master:ker} as special cases
when one selects $\Vker$, $\arFn := (\arFn_0, \ldots, \arFn_p)$ and
$S := (S_0, \ldots, S_p)$ appropriately.  The
reader may refer to \cite{glatt2024parallel} for a more detailed and comprehensive account.
In particular, Theorem 2.2 of \cite{glatt2024parallel}
provides the appropriate constraints, and we reproduce this result
here.  To do so, we need the notion of the pushforward
of a probability measure $\nu$ under a map $F$.  Suppose that
$(Z, \Sigma_Z)$ and $(Z', \Sigma_{Z'})$ are measurable spaces and that
$F : Z \to Z'$ is measurable.  Then, given a probability measure $\nu$
on $(Z, \Sigma_Z)$, $F^*\nu$, referred to as the pushforward to $\nu$
under $F$ is defined as
\begin{align}\label{eq:push:for:def}
F^*\nu(A) = \nu(F^{-1}(A))
\end{align}
for $A \in \Sigma_{Z'}$.  At the level of random variables, if
$\xi \sim \nu$, then $F(\xi) \sim F^*\nu$.  When $Z = \RR^N$ and
when $\nu$ is continuously distributed so that
$\nu(d \bq) = \dnu(\bq) d\bq$, then, for any $C^1$ diffeomorphism $F: \RR^N
\to \RR^N$,
\begin{align}\label{eq:push:cont:case}
  F^* \nu(d\bq) = \dnu(F^{-1}(\bq)) | \det \nabla F^{-1}(\bq)| d\bq.
\end{align}  
With this preliminary in hand, criteria for the reversibility of
\cref{alg:main:master} are as follows.
\begin{theorem}\label{thm:mprop}
  Fix $p \geq 1$ and take
  \begin{align}\label{eq:comp:measure}
    \cM(d \bq, d\bv) := \Vker (\bq, d \bv)\mu(d \bq).
  \end{align}  
  Suppose that
  \begin{enumerate}[label=(\roman*)]
  \item\label{P1} $S_j$ is an involution, i.e. $S_j \circ S_j = I$, for $j = 0, 1,\ldots,p$;
  \item\label{P2} For every
    $j=0,1,\ldots, p$,
     $\arFn_j( S_j(\bq, \bv)) S^*_j \cM(d \bq, d \bv)
    = \arFn_j(\bq, \bv) \cM(d \bq,d \bv)$.
  \end{enumerate}
  Then, under \ref{P1} and \ref{P2}, \cref{alg:main:master} is
  unbiased with respect to $\mu$.  More precisely, under these
  conditions, $\Pker$ defined as \eqref{eq:main:master:ker} is
  reversible with respect to $\mu$ namely
  $\Pker(\bq,d\btq) \mu(d\bq) = \Pker(\btq,d\bq) \mu(d\btq)$ and hence
  $\mu$ is invariant under $P$.
\end{theorem}

The proof of \cref{thm:mprop}, found in \cite{glatt2024parallel}, is a
straightforward computation using a calculus for push-forward maps \eqref{eq:push:for:def}. The depth
of this result is in finding a simple formulation that
encompasses so many known and yet to be discovered algorithms in a
useful way as we unfold below.  

We make a few notes before proceeding to these applications.
First, we note that
the map from $(\Vker, S, \arFn)$ to $P$ is
 not one-to-one; multiple different choices can sometimes yield the
same algorithm as demonstrated in \cite[Section 3]{glatt2023accept}.
Second, under the further assumption that
\begin{align}\label{eq:abs:cont}
  \text{$S_j^* \cM$ is absolutely continuous with
  respect to $\cM$},
\end{align}  
the condition in \ref{P2} reduces to
\begin{align}\label{eq:ar:RND:form}
  \arFn_j( S_j(\bq, \bv)) \frac{d S^*_j \cM}{d \cM}(\bq, \bv)
    = \arFn_j(\bq, \bv),
\end{align}
where $\frac{d S^*_j \cM}{d \cM}$ is the Radon-Nikodym derivative.
Note that, as observed in \cite[Corollary
2.6]{glatt2024parallel}, in the case when $S_0$ is the identity, we can take
\begin{align}\label{eq:ar:RND:form:MH}
  \arFn_j(\bq, \bv) \propto \frac{d S^*_j \cM}{d \cM}(\bq, \bv) \wedge 1, \,\, j=1, \ldots, p; \quad \arFn_0 \equiv 1 - \sum_{j=1}^p \arFn_j
\end{align}
to derive the classical Metropolis-Hastings ratio in
\cref{sec:MH:classic}.  While 
\cite{Tierney1998} shows that \eqref{eq:ar:RND:form:MH} is the
optimal choice in terms of Peskun's ordering \cite{peskun1973optimum} for the single proposal $p =1$ case,
\eqref{eq:ar:RND:form} accommodates other choices such
as the Barker ratio \cite{barker1965monte}, which regains salience in the multiproposal
context $p \gg 1$; see \eqref{eq:multi:prop:tj:3}
and \cref{alg:mpCN} below. Furthermore, in the special case of
continuous target distributions, \eqref{eq:push:cont:case} with
\eqref{eq:ar:RND:form} allows us to derive explicit formulas for
$\arFn$ given $S_j$ and $\Vker$.  The Cameron-Martin theorem plays
an analogous role when we address Hilbert space methods in
\cref{sec:Hilbert}.

One might ask in general why it is worth the effort to develop
\cref{alg:main:master} and \cref{thm:mprop} at the level of general
measurable spaces.  One immediate answer is that this approach allows
for the treatment of discrete and continuous target measures at once and,
by extension, allows one to consider all the intermediate cases that sometimes arise.  Second, we are often interested in infinite-dimensional
target measures defined over spaces of
functions (\cref{sec:Hilbert}). 

\subsection{Connection to
  Metropolis-Hastings and Green's Jacobian Extension}
\label{sec:MH:classic}

One can use \cref{thm:mprop} to derive the classical Metropolis-Hastings algorithm, and we do so under the general state space
formulation of \cite{green1995reversible} and \cite{Tierney1998}.  Consider a
measurable space $(X, \Sigma_X)$, a target measure $\mu$ on $X$,
take $Y = X$ and consider
any proposal kernel $\Vker: X \times \Sigma_X \to [0,1]$.  Take
$p =1$ and select the involutive mappings
\begin{align}\label{eq:flip:S:classic}
  S_0(\bq, \bv) = (\bq,\bv),
  \quad
  S_1(\bq, \bv) = (\bv, \bq).
\end{align}  
Criterion \ref{P2} is satisfied trivially for $j = 0$ and any
$\arFn_0$, while for $j = 1$, \ref{P2} reduces to the requirement that
\begin{align}\label{eq:p2:cond:satisfied}
  \arFn_1(\bv, \bq) \cM(d\bv, d\bq) =   \arFn_1(\bq, \bv)\cM(d\bq, d\bv).
\end{align}
Thus, under the absolute continuity assumption \eqref{eq:abs:cont}, we can select
\begin{align}\label{eq:tierney:gen}
  \arFn_1(\bq, \bv) := \frac{d S_1^* \cM}{d \cM}(\bq, \bv) \wedge 1,
  \quad \arFn_0(\bq, \bv) = 1 - \arFn_1(\bq, \bv),
\end{align}
in order to satisfy \eqref{eq:p2:cond:satisfied}.
Note that, when $X = \RR^N$ and we are in the classical
continuous target distribution case
\begin{align}\label{eq:cont:MH:class:1}
  \mu(d\bq) = \dmu(\bq) d \bq,
  \quad
  \Vker(\bq, d\bv) = \dVker(\bq, \bv) d\bv,
\end{align}  
then $\arFn_1$ reduces
to the Metropolis-Hastings ratio
\begin{align}\label{eq:cont:MH:class:2}
  \arFn_1(\bq, \bv) = \frac{\dmu(\bv)\dVker(\bv, \bq)}{\dmu(\bq)\dVker(\bq, \bv)} \wedge 1
\end{align}
according to \eqref{eq:push:cont:case}.
Hence, \cref{alg:main:master} reduces to the classical Metropolis-Hastings
algorithm under \eqref{eq:flip:S:classic}, \eqref{eq:cont:MH:class:1} and
\eqref{eq:cont:MH:class:2}.

Under a slight variation, we have the following extension which recovers the main result in
\cite[Theorem 2.1]{glatt2023accept} from \cref{thm:mprop},
while also yielding 
a general state space extension of the MHGJ algorithm.  To obtain this
result, we suppose
now that $(Y, \Sigma_Y)$ is an arbitrary state space with $\Vker$ any
kernel on $X \times \Sigma_Y$.  Take
$S: X \times Y \to X \times Y$ to be any (measurable) involution so
that $S^* \mathcal{M}$ is absolutely continuous with respect to $\cM$.
With this $\Vker$ and $S$, we now take $S_0$ as the identity and
$S_1 = S$ to supply the two involutions.  Relative to these more general definitions,
we take $\arFn_0, \arFn_1$ as in \eqref{eq:tierney:gen} to derive that
\begin{align}
  \Pker (\bq, d \btq) 
  = \int_\spv &\left(\frac{d S^* \cM}{d \cM}(\bq, \bv) \wedge 1 \right)
     \delta_{\Pi_1 \circ S(\bq,\bv)}(d\btq) \Vker(\bq, d \bv) \notag\\
              &\qquad + \delta_\bq(d\btq)
                \int_\spv \left(1 - \frac{d S^* \cM}{d \cM}(\bq, \bv) \wedge 1\right)
     \Vker(\bq, d \bv)
    \label{eq:gen:Pker:single}
\end{align}
is reversible with respect to $\mu$.   This corollary 
of \cref{thm:mprop} is sufficient to encompass many reversible, single-proposal MCMC algorithms,
including the cases outlined in \cref{sec:Ham:Methods} and
\cref{sec:Hilbert}.
To see that we have produced a `general state space' extension of MHGJ
we simply take $X = \RR^N$ and consider a continuous target $\mu$ as
in \eqref{eq:cont:MH:class:1}.  Taking now $Y = \RR^M$, any continuous
kernel $\Vker(\bq, d\bv) = \dVker(\bq, \bv) d\bv$ and any $C^1$ involution
$S: \RR^{N+M} \to \RR^{N+M}$ (and noting
\eqref{eq:push:cont:case} and that $S = S^{-1}$), the acceptance probability becomes
\begin{align}\label{eq:MHGJ:AR}
  \frac{d S^* \cM}{d \cM}(\bq, \bv) \wedge 1 =
  \frac{g(S(\bq,\bv)) |\det \nabla S(\bq,\bv)|} {g(\bq,\bv)} \wedge 1,
  \quad \text{ where } g(\bq, \bv) := \dmu(\bq) \dVker(\bq, \bv).
\end{align}
This is precisely the MHGJ acceptance probability from
\cite{green1995reversible}.

\subsection{Hamiltonian Methods}
\label{sec:Ham:Methods}

\newcommand{\mass}{\mathbf{M}}

The standard HMC algorithm \cite{duane1987hybrid,neal2011mcmc} samples from a continuous target distribution $\mu$ on
$\RR^N$, typically written in the convenient form
$\mu(d\bq) \propto e^{-\Pot(\bq)} d \bq$, for some `potential'
function $\Pot: \RR^N \to \RR$. In this setting, one generates proposals using the Hamiltonian dynamic
\begin{align}\label{fd:ham:dyn}
  \frac{d \bq}{dt} = \mass^{-1}\bv, \quad
  \frac{d \bv}{dt} = - \nabla \Pot(\bq),
\end{align}
on the extended phase space $\RR^N \times \RR^N$, for some
symmetric positive definite `mass matrix' $\mass \in \RR^{N \times N}$, so
that we have the associated Hamiltonian function
\begin{align}\label{eq:Ill:have:the:classic:ham}
  \Ham(\bq, \bv) := H_1(\bq, \bv) + H_2(\bq,\bv) =\Pot(\bq)
  + \frac{1}{2} \langle \mass^{-1}\bv,\bv\rangle.
\end{align}
Here, recall that the Gibbs measure,
\begin{align}
  \cM(d \bq, d \bv) \propto \exp( - \Ham(\bq, \bv) ) d \bq d \bv,
\end{align}  
is invariant under the dynamic $\hat{S}_T$ defined by
\eqref{fd:ham:dyn}, i.e., $\hat{S}_T^* \cM = \cM$, for every $T > 0$.
Thus, from a current state $\bq^{(k)}$, drawing from the momentum
marginal of $\cM(d \bq, d \bv)$, namely taking
$\bv^{(k)} \sim \cN (\boldsymbol{0},\mass)(d\bv)$, and setting
$\bq^{(k+1)} :=\Pi_1 \hat{S}_T(\bq^{(k)}, \bv^{(k)})$, we obtain a
Markov chain $\{\bq^{(k)}\}_{k \geq 0}$ that is stationary with
respect to $\mu$. 

Of course, one can rarely resolve the map $\hat{S}_T$ analytically, so one uses a numerical
approximation to \eqref{fd:ham:dyn} in practice. Here, a
common choice is the classical \emph{leapfrog} or \emph{Verlet}
integrator, which one derives by splitting \eqref{fd:ham:dyn} into the
dynamics associated with $H_1$ and $H_2$ namely $d\bq/dt = 0$,
$d\bv/dt = - \nabla \Pot(\bq)$, and $d\bq/dt = \mass^{-1}\bv$,
$d\bv/dt = 0$.  These systems admit the closed form
solutions
\begin{align}\label{eq:class:split}
  \Xi^{(1)}_t(\bq,\bv) = (\bq, \bv - t \nabla \Pot(\bq)),
  \quad \text{ and } \quad
  \Xi^{(2)}_t(\bq,\bv) = (\bq + t \mass^{-1}\bv, \bv),
\end{align}
respectively.  We now take
\begin{align}\label{eq:HMC:spec}
  \Vker(\bq,d\bv) = \cN(\boldsymbol{0},\mass)(d\bv),
  \quad
  S(\bq,\bv) = R \circ (\Xi^{(1)}_{\dt/2} \circ \Xi^{(2)}_{\dt} \circ
  \Xi^{(1)}_{\dt/2})^n,
\end{align}  
where $R(\bq,\bv) = (\bq, -\bv)$ is the momentum-flip
operator. Here, $\dt > 0$ is the time step size, $n$ is the number of
leapfrog steps, and $T:= n\dt$ is the total
integration time of the approximate trajectory.

It is clear by direct computation that, for any $t > 0$, both
$R \circ \Xi^{(1)}_{t}$ and $R \circ \Xi^{(2)}_{t}$ are involutions.
Furthermore, each of $R$, $ \Xi^{(1)}_{t}$ and $ \Xi^{(2)}_{t}$ preserve
volumes.  With these properties, and in view of the palindromic
structure of the Verlet integrator in \eqref{eq:HMC:spec}, we see that
$S$ is itself involutive and volume preserving.  Thus, under these
definitions for $S$ and following \eqref{eq:push:cont:case}, we have that
\begin{align}\label{eq:HMC:class:AR}
  \frac{d S^*\cM}{d \cM }(\bq,\bv)
  = \exp(\Ham(\bq,\bv) - \Ham(S(\bq,\bv)) .
\end{align}
In particular, we can see the HMC algorithm falls out as a special case
of MHGJ by taking $g(\bq,\bv) \propto \exp(-H(\bq,\bv))$ in
\eqref{eq:MHGJ:AR}. By extension, we see that HMC is a special case of
\cref{alg:main:master} under \cref{thm:mprop}.

The many extensions to classical HMC all fall under the aegis of \cref{thm:mprop}, thus underlining the connection with the MHGJ algorithm.  This family of
Hamiltonian methods includes the No-U-Turn sampler
\cite{hoffman2014no}, Riemannian manifold HMC
\cite{girolami2011riemann}, $\infty$HMC
\cite{beskos2011hybrid} (Section \ref{sec:Hilbert}), and a variety of
other algorithms advanced in, e.g., \cite{neal2011mcmc}, \cite{bou2018geometric} and \cite{glatt2023accept}.  As laid out in detail in
\cite{andrieu2020general}, \cite{glatt2023accept}, \cite{neklyudov2020involutive} and \cite{glatt2024parallel}, the combination of \cref{alg:main:master} with
\cref{thm:mprop} acts as a sort of skeleton key to navigate,
classify and justify this zoo of different species of extended phase
space methods.  Here, note that the involutive condition in
\cref{thm:mprop} and the Jacobian term in \eqref{eq:MHGJ:AR} emphasize the
indispensable role of path-reversibility and volume preservation in the
determination of a reasonable acceptance ratio.

\subsection{Surrogate-Trajectory HMC}

A computational bottleneck
associated with the HMC methodology is the calculation of gradients
within the numerical integration of the Hamiltonian trajectory
as in, e.g., \eqref{eq:HMC:spec}.  Thus, an important question is how one may replace such
gradient terms with an approximation or `surrogate' that helps overcome the random-walk behavior of Metropolis-Hastings-type samplers even if failing to provide an accurate approximation to the Hamiltonian dynamic \eqref{fd:ham:dyn}.  In this framing,
\cref{thm:mprop} provides justification for a variety of
possible surrogate-based approaches.
While \cite{glatt2023accept} lay out many
 possibilities for surrogates, we detail just one case here and refer to
 this work for a number of extensions. 

We proceed by approximating \eqref{fd:ham:dyn} with the dynamic
$d\bq/dt = f_1(\bv), d\bv/dt = f_2(\bq)$ and integrate by splitting
into position and momentum updates, extending \eqref{eq:class:split} to
define
\begin{align}\label{eq:sj:split}
  \Xi^{(1)}_t(\bq,\bv) = (\bq, \bv - t f_2(\bq)),
  \quad \text{ and } \quad
  \Xi^{(2)}_t(\bq,\bv) = (\bq + t f_1(\bv), \bv).
\end{align}
Relative to these new definitions we then consider the same leapfrog
integrator \eqref{eq:HMC:spec} to define $S$.  Under the mild
assumption that $f_1$ is an odd function, namely that
$f_1(\bv) = - f_1(-\bv)$ for $\bv \in \RR^\dimx$, it follows that $S$
is an involution. Volume preservation also follows so long as
$f_1, f_2$ are $C^1$.  Hence, we may
use the usual acceptance ratio \eqref{eq:HMC:class:AR} as in the case of
classical HMC.
From the point of view of reversibility, we can replace
$\nabla \Pot$ with any (sufficiently smooth) approximation.  Note
furthermore that, to accommodate a state-dependent mass matrix
we can take
$\Vker(\bq, d \bv) := Z_{\mathcal{K}}(\bq)^{-1} e^{-\mathcal{K}(\bq,\bv)}
d\bv$ where
$Z_{\mathcal{K}}(\bq) = \int_{\RR^N}e^{-\mathcal{K}(\bq,\bv)} d\bv$
and modify the definition of $H$ to
$H(\bq,\bv) = \Phi(\bq) + \mathcal{K}(\bq,\bv) + \ln
Z_{\mathcal{K}}(\bq)$ to determine our acceptance ratio exactly as in
\eqref{eq:HMC:class:AR}.  Here, as in
\cite{beskos2017geometric}, \eqref{eq:sj:split} would then represent a
separable approximation of a non-separable dynamic.  Per \cite{glatt2023accept}, one may also develop a surrogate version of the implicit
scheme in \cite{girolami2011riemann}.

A small literature on surrogate-trajectory HMC methods precedes the arrival
of the complete involutive theory of MCMC yet highlights possible approaches to the
selection of approximate dynamics as in, e.g., $f_1, f_2$ of
\eqref{eq:sj:split}.  To summarize this literature at a high level,
there are two general ways to proceed. First, one may consider a
`surrogate model'-based approach by approximating $\Phi$ or $\nabla \Phi$
with some kind of a surrogate statistical model such as a Gaussian process (GP), a neural
network, a piecewise-constant function, etc. Alternatively, one may
take advantage of some specific structure of $\nabla \Phi$ to derive
an analytical approximation by, e.g., truncating an
appropriate series expansion or otherwise leveraging the target model structure.

Within the surrogate model approach,
\cite{rasmussen2003gaussian} approximates the log-posterior
distribution using GP regression on a set of
samples generated by vanilla HMC during an exploratory
phase. Differentiating the GP's conditional mean function results in
an approximation to the gradient.  The author applies this GPHMC to a
few toy problems, but correctly notes that the method ``allows
Bayesian treatment of models with posteriors that are computationally
demanding, such as models involving computer simulation.''  In the
context of Bayesian inverse problems (\cref{sec:BIP}),
\cite{lan2016emulation} use GPs to approximate log-posterior
gradients and Hessians, relying on statistical emulation theory to
choose design points on which to condition. \cite{zhang2017precomputing} precompute gradients on a sparse
grid and use piecewise-constant interpolations of log-posterior
gradients to construct surrogate-trajectories for a range of inference
problems.  Next, \cite{zhang2017hamiltonian,zhang2018variational} use
shallow neural networks to approximate the log-posterior itself,
letting the gradient of the neural network approximate the
log-posterior gradient.  Building on this work, \cite{li2019neural}
use a neural network to directly approximate the targeted gradient
itself and demonstrate performance gains over the preceding indirect
log-posterior approximations for a range of problems.  
Section \ref{sec:BIP} presents a neural network-based
surrogate-trajectory version of the $\infty$HMC (\cref{sec:Hilbert}) for Bayesian inversion of nonlinear PDEs.
Here, challenging nonlinear and multimodal target geometry causes
difficulties for surrogate-trajectory approaches that nonetheless
outperform classical HMC.

Within the analytical approach,
\cite{magee2024random} and \cite{didier2024surprising} consider phylogenetic continuous time
Markov chain  models for which the log-likelihood
gradient admits a particular series representation. Naively truncating
this series to include a single term reduces computing time and boosts
performance.  We discuss this example in Section \ref{sec:ctmc}. Section \ref{sec:bmds} provides a negative example in which the analytic approach fails to speed up inference. This example involves truncating the Bayesian
multidimensional scaling log-posterior gradient for probabilistic
dimension reduction.  Here, the gradient for a low-dimensional
embedding of a single data point relies on the remaining $N-1$
observed data.  Using a small subset the exact gradient's $N (N-1)$
terms within surrogate-trajectory HMC fails to accelerate inference.

\subsection{Hilbert Space Methods}
\label{sec:Hilbert}

\newcommand{\cJ}{\mathcal{J}}

We turn to describe the Hilbert space methods established in \cite{neal1999regression}, \cite{beskos2008mcmc}, 
  \cite{beskos2011hybrid} and \cite{cotter2013mcmc}.  These algorithms provide an effective means of sampling certain classes of infinite-dimensional measures arising in, e.g., Bayesian inverse problems in which the unknown is a function specified by infinitely many parameters (\cref{sec:BIP}).  More precisely, these methods address
measures $\mu$ defined on a separable Hilbert space $X$ where $\mu$ is absolutely continuous with respect to a Gaussian measure
$\mu_0 = \cN(\boldsymbol{0},\cC)$ on $X$, for some covariance operator
$\cC: \spq \to \spq$ that is trace-class, symmetric and strictly
positive definite.  Such target measures $\mu$ take the
form
\begin{align}\label{eq:target:Hilbert}
  \mu(d\bq) \propto
  \exp\left(-\Pot(\bq) -
  \frac{1}{2} \langle \cC^{-1} \bq, \bq \rangle \right) d \bq,
\end{align}
where $\Phi: X \to \RR$ corresponds to the log-likelihood in a
typical Bayesian setting.  This expression for $\mu$ is only formal at the infinite-dimensional level where there is no notion of Lebesgue measure.  We
refer the reader to, e.g., \cite{da2014stochastic} or
\cite{hairer2009introduction} for a detailed treatment of Gaussian
measures on function space.

The three original Hilbert space methods, pCN, $\infty$MALA and
$\infty$HMC, represent infinite-dimensional generalizations of random-walk Metropolis, Metropolis-adjusted Langevin algorithm (MALA) and HMC, respectively.  The first, pCN, stands for
`preconditioned Crank-Nicolson': the proposal in this method represents a semi-implicit Crank-Nicolson discretization of an
Ornstein-Uhlenbeck (OU) process preconditioned by the covariance
operator $\cC$ defining $\mu_0$.  Such a proposal structure leaves
$\mu_0$ invariant.  Thus, under the choice of $S_0, S_1$ in
\eqref{eq:flip:S:classic} and following \eqref{eq:tierney:gen}, we have the
proposal and acceptance probabilities 
\begin{align}\label{eq:pCN:summary}
  \Vker(\bq, d \bv) = \cN(\rho \bq, (1 - \rho^2) \cC)( d\bv),
  \quad \arFn_1(\bq,\bv) = \exp(\Phi(\bq) - \Phi(\bv)) \wedge 1,
\end{align}
respectively.  Here $\rho \in [0,1]$ is a parameter, related to the
size of the time step in the OU discretization, that
determines how far proposals tend to be from a current state.
To incorporate some of the local structure of $\Phi$ in
\eqref{eq:target:Hilbert} at each step,  $\infty$MALA 
\cite{cotter2013mcmc} considers proposals from the
preconditioned Langevin stochastic dynamic
$d \bq + \frac{1}{2}(\bq + \cC D \Phi(\bq))dt = \cC^{1/2}dW$.  This
method again falls out of \cref{eq:main:master:ker} and \cref{thm:mprop}
in several different ways \cite{glatt2023accept}.

\begin{algorithm}[!t]
\caption{\cite{glatt2023accept}}
\begin{algorithmic}[1]\label{alg:inf:surgate}
  \State  Select the algorithm parameters:
  \begin{itemize}
  \item[(i)] The proposal kernel with local preconditioning
    $\Vker(\bq,d \bv) := \exp(-\Psi(\bq, \bv))\mu_0(d\bv)$.
  \item[(ii)] The surrogate $\agp$ for $\cC D \Pot$.
  \item[(iii)] Relative step sizes $\da > 0$, $\db >
    0$ and number of steps $n$.
  \end{itemize}  
    \State Choose $\bq^{(0)} \in \spq$.
    \For{$k \geq 0$}
    \State Sample $\bv^{(k)} \sim \Vker(\bq^{(k)}, d \bv)$.
    \State Set $(\bq^{(k)}_0, \bv^{(k)}_0) = (\bq^{(k)}, \bv^{(k)})$.
    \For{$l = 1, \ldots, n$}
    \State Set $(\bq^{(k)}_l,\bv^{(k)}_l) = \Xi_{\da}^{(1)} \circ
    \Xi_{\db}^{(2)}\circ\Xi_{\da}^{(1)}(\bq^{(k)}_{l-1},\bv^{(k)}_{l-1})$.
    \EndFor
    \State Compute
    \begin{align}
      \! \!\! \! \! \!\arFn_k &=
                \exp \left( \Pot(\bq_0^{(k)}) + \Psi(\bq_0^{(k)}, \bv_0^{(k)})
                        - \Pot(\bq_n^{(k)}) -\Psi(\bq_n^{(k)}, -\bv_n^{(k)})
                           \right.
                           \notag\\
              &\qquad \qquad
                + 2 \da \sum_{i=1}^{n-1}
           \langle \cC^{-1/2}\bv_i^{(k)}, \cC^{-1/2}f(\bq_i^{(k)}) \rangle -\frac{\da^2}{2} \left[ |\cC^{-1/2} f(\bq_0^{(k)})|^2
                - |\cC^{-1/2} f(\bq_n^{(k)})|^2 \right]
      \notag\\
		&\qquad \qquad \left.
           + \da  \left[ \langle \cC^{-1/2} \bv_0^{(k)}, \cC^{-1/2}f(\bq_0^{(k)}) \rangle
           + \langle \cC^{-1/2} \bv_n^{(k)}, \cC^{-1/2}f(\bq_n^{(k)}) \rangle \right] \right),
           \label{eq:RN:SM:M}
\end{align}

        \State Set $\bq^{(k+1)} =\bq^{(k)}_n$ with probability
                   $\arFn_k \wedge 1$,
                   otherwise set $\bq^{(k+1)} = \bq^{(k)}$.
    \EndFor
\end{algorithmic}
\end{algorithm}

 \cite{beskos2011hybrid} introduces $\infty$HMC, the
infinite-dimensional version of the HMC. Here, to sample
from a target $\mu$ as in \eqref{eq:target:Hilbert}, we consider
a Hamiltonian of the form $\Ham := \Ham_1 + \Ham_2$, where
\begin{align}\label{eq:infHMC:Ham}
  \Ham_1(\bq,\bv) := \frac{1}{2} |\cC^{-1/2}\bv|^2
  + \frac{1}{2}|\cC^{-1/2} \bq|^2,\quad
  \Ham_2(\bq,\bv) := \Pot(\bq).
\end{align}
The non-canonical, position-velocity formulation of the corresponding Hamiltonian dynamic becomes
\begin{align}\label{eq:infHMC:dym}
  \frac{d\bq}{dt} = \bv,
  \quad
  \frac{d\bv}{dt} = - \bq - \cC D \Pot(\bq).
\end{align}
This equation falls out of \eqref{fd:ham:dyn} above by taking
$\mass = \cC^{-1}$ and replacing the momentum variable $\bv$ in
\eqref{fd:ham:dyn} with velocity via the change of variable
$\cC \bv \rightsquigarrow \bv$.\footnote{In more general terms, we may
  consider the invertible, anti-symmetric operator $\cJ$ in the
  associated Hamiltonian system $d\bz/dt = \cJ^{-1} D \Ham(\bz)$.  In
  this case,
  $\cJ = \begin{bmatrix} 0 & - \cC^{-1} \\ \cC^{-1} & 0 \end{bmatrix}$
  yields \eqref{eq:infHMC:dym}.}  At each step, one redraws the velocity
component from $\Vker(\bq, d \bv) = \mu_0(d \bv)$ so that
our Gibbs measure takes the form
$\cM(d \bq, d \bv) \propto \exp(- \Phi(\bq)) \mu_0(d \bq) \mu_0(d
\bv)$.

To numerically integrate \eqref{eq:infHMC:dym}, we split the dynamics
according to $\Ham_1, \Ham_2$ in \eqref{eq:infHMC:Ham} and obtain exact solutions
\begin{align}\label{eq:infty:HMC:step}
  \!\!\Xi_t^{(1)}(\bq,\bv) = (\bq, \bv - t \cC D \Pot(\bq)),
  \quad
\Xi_t^{(2)}(\bq,\bv) = (\bq \cos t + \bv \sin t,\bv \cos t - \bq \sin t ).
\end{align}
Then, once again using the palindromic (leapfrog) splitting in 
\eqref{eq:HMC:spec} to define $S:X \times X \to X \times X$, this time
with the steps specified by \eqref{eq:infty:HMC:step}, we obtain an
involution.  On the other hand, it is not clear what volume
preservation means at the infinite dimensional level.  That said,
letting $\cM_0 := \mu_0 \otimes \mu_0$ results in
$(\Xi_t^{(2)})^* \cM_0 = \cM_0$ and $\Xi_t^{(1)}\cM_0$ being
absolutely continuous with respect to $\cM_0$ due to the Cameron-Martin theorem (see, e.g.,
\cite{da2014stochastic}). Thus, we can show that $S^* \cM$ is absolutely
continuous with respect to $\cM$ and compute
an explicit expression for $d S^* \cM/d\cM$; see \eqref{eq:RN:SM:M}.

Thus, we can explicitly show that $\infty$HMC is a special case of the
generalized MHGJ given by \eqref{eq:gen:Pker:single} \cite[Section 4]{glatt2023accept}.   This approach provides
an alternative to the original derivation of the acceptance
probability in $\infty$HMC established via an involved finite
dimensionalization procedure in \cite{beskos2011hybrid}.  Moreover,
this generalized MHGJ approach yields a wider class of novel surrogate
methods in which one replaces the potentially computationally intensive terms
$\cC D \Pot(\bq)$ in \eqref{eq:infty:HMC:step} with
a general approximation $f$. \cref{alg:inf:surgate} summarizes this new class of surrogate methods.
 We note that pCN, $\infty$MALA and $\infty$HMC all fall out as special cases of
\cref{alg:inf:surgate} under specific parameter choices
\cite[Section 5]{glatt2023accept}. Thus, \cref{thm:mprop} leading to
\cref{alg:inf:surgate} illustrates the unifying potential of an involutive theory that simultaneously provides opportunity for novel method development.

\subsection{Multiproposal Methods}
\label{sec:Multiprop}

While HMC tends to perform better on a per sample basis than more
traditional random-walk approaches, its benefits may wane when one
accounts for the cost of computing gradients. These high costs provide
motivation for the surrogate approaches discussed above, but finding
an appropriate surrogate is often a bespoke challenge.  Moreover,
gradient-based approaches often fail to sample efficiently from
multimodal targets.  A multiproposal approach may provide a viable
alternative strategy in these contexts.  When acceptance
probabilities take a Barker-like, target-informed structure, this
multiproposal algorithm can leverage a parallel computing strategy different from, and complementary to, the embarrassingly parallel
approach of running multiple chains \cite[Section 5.1]{glatt2024parallel}.

Multiproposal MCMC methods feature across a variety of recent
works \cite{neal2003markov, tjelmeland2004using,
  frenkel2004speed, delmas2009does, calderhead2014general,
  luo2019multiple, liu2000multiple, schwedes2021rao,
  Holbrook2023generating,holbrook2023quantum}. The generalization recalled in \cref{alg:main:master} and \cref{thm:mprop} provides a unified foundation for this seemingly disparate literature.  This involutive approach once again facilitates the
derivation of new methods such as the multiproposal Hilbert space method given by \cref{alg:mpCN}.

One approach to pulling interesting methods out of
\cref{alg:main:master} and \cref{thm:mprop} draws inspiration from the
conditionally independent proposal structure found in
\cite{tjelmeland2004using}.  This approach proceeds by drawing a
single point $\bar{\bq}^{(k)}$ around the current state
$\bq^{(k-1)}:= \bq_{0}^{(k)}$ with distribution
$\bar{Q}(\bq^{(k)}_0, d \tilde{\bq})$ and then producing a cloud
of proposal points $\bq^{(k)}_1, \ldots, \bq^{(k)}_p$
around $\bar{\bq}^{(k)}$ independently with distributions
$Q(\bar{\bq}^{(k)}, d \tilde{\bq})$.  To finish the connection to
\cref{alg:main:master}, we take
\begin{align}
  &\bv = (\bq_1, \ldots, \bq_p) \in Y := X^p,
  \quad
  \Vker(\bq_0, d \bq_1, \ldots, d \bq_p)
    := \int_X \Pi_{i =1}^p Q(\tilde{\bq}, d \bq_i) \bar{Q}(\bq_0, d \tilde{\bq}),
  \label{eq:multi:prop:tj:1}\\
  &S_0 = I, \quad S_j(\bq_0,\bq_1, \ldots, \bq_p)
    = ( \bq_j, \bq_1, \ldots, \bq_{j -1}, \bq_0, \bq_{j+1}, \ldots, \bq_p),
    \text{ for } j = 1, \ldots, p,
      \label{eq:multi:prop:tj:2}
\end{align}  
where $X^p$ is the $p$-fold product of $X$.  To obtain a desirable
class of Barker-like or target-informed acceptance ratios, we
assume we can find a $\sigma$-finite measure $\mu_0$ such that the
target measure $\mu$ and $\mu_0$ are mutually absolutely continuous.
We then suppose that the balance-like condition
\begin{align}\label{eq:gen:bal:cond}
  Q(\bq, d \tilde{\bq})\mu_0(d \bq)
  = \bar{Q}(\tilde{\bq}, d\bq)\mu_0(d \tilde{\bq})
\end{align}  
holds.  Under this assumption,
one may show that, by taking 
\begin{align}
  \arFn_j(\bq_0, \ldots, \bq_p)
  := \frac{\frac{d \mu}{d \mu_0}(\bq_j)}
  {\sum_{k = 0}^p \frac{d \mu}{d \mu_0}(\bq_k)} ,
          \label{eq:multi:prop:tj:3}
\end{align}
the algorithm corresponding to $(\Vker, S, \arFn)$ from
\eqref{eq:multi:prop:tj:1}, \eqref{eq:multi:prop:tj:2} and
\eqref{eq:multi:prop:tj:3} holds $\mu$ in detailed balance \cite[Corollary 2.12]{glatt2024parallel}.
Out of \eqref{eq:multi:prop:tj:1}, \eqref{eq:multi:prop:tj:2} and
\eqref{eq:multi:prop:tj:3} falls the original algorithm of
\cite{tjelmeland2004using} with the Gaussian proposal kernel condition
removed. In fact, one may obtain a more general formulation that allows for
a Hastings-like asymmetric proposal structure \cite[Algorithm
4.1]{glatt2024parallel}.  This setup also produces a
multiproposal generalization of the pCN algorithm by taking
$Q = \bar{Q}$ as the pCN kernel given by \eqref{eq:pCN:summary} (\cref{alg:mpCN}).

 \begin{algorithm}[!t]
\caption{Multiproposal pCN to sample $\mu(\bq) \propto e^{ - \Phi(\bq)} \mu_0(d\bq)$ with $\mu_0 = \cN(0,\cC)$}
\begin{algorithmic}[1]
    \State Select the algorithmic parameter $\rho \in [0,1]$ and the number of proposals $p \geq 1$.
    \State Choose $\bq^{(0)} \in \spq$.
    \For{$k \geq 1$}
        \State Draw $\bar{\bq}^{(k)}=  \rho \bq^{(k-1)}+ \sqrt{1 -\rho^2} \xi^{(k)}$, $\xi^{(k)} \sim \cN(0,\cC)$. 
        \State Set $\bq_{0}^{(k)} = \bq^{(k-1)}$ and take $\bq_{j}^{(k)} =  \rho\bar{\bq}^{(k)}+ \sqrt{1 -\rho^2} \xi_{j}^{(k)}$,
        by drawing i.i.d. $\xi_{j}^{(k)} \sim \cN(0,\cC)$, for $j=1, \ldots, p$.
        \State Set $\bq^{(k)} =\bq_j^{(k)}$ with probabilities:
        \begin{align*}
	   \arFn_j(\bq_{0}^{(k)}, \ldots, \bq_{p}^{(k)} ) = \frac{\exp( - \Phi(\bq_{j}^{(k)} ))}{\sum_{l =0}^p \exp( - \Phi(\bq_{l}^{(k)} ))}, \quad j=0, \ldots, p.
	\end{align*}
    \EndFor
\end{algorithmic}\label{alg:mpCN}
\end{algorithm}

The conditionally independent structure is not the only
multiproposal algorithm that one may derive from the framework of
\cref{thm:mprop}.  Other such algorithms include several multiproposal
versions of the algorithms discussed in the examples above, such as
Hamiltonian-type methods; for example, multiproposal structure
arises from the different steps along a Hamiltonian integration path \cite[Algorithm
4.1]{glatt2024parallel}.
Additionally, the framework around \cref{thm:mprop} further
extends to account for algorithms that at each iterative step select
multiple states out of the cloud of proposals, yielding a
generalization of schemes from \cite{calderhead2014general}.

\section{Case Studies in Surrogate-Trajectory HMC}\label{sec:surrog}

While the involutive theory provides an elegant framework for a
general class of surrogate-trajectory HMC methods, the exact form
taken by such methods depends on the problem at hand,
the models used and the training and creativity of the practitioner.
Surrogate-trajectory HMC, in its current form, is more art than
science. In the following, we provide a few illustrations showing how
surrogate-trajectory methods can look in the wild and conclude with a
list of open questions that stand in the way of surrogate-trajectory
methods' broader uptake.

\subsection{Continuous-Time Markov Chains}\label{sec:ctmc}

\newcommand{\QQ}{\mathbf{Q}}
\newcommand{\ppi}{\boldsymbol{\pi}}
\newcommand{\ttheta}{\boldsymbol{\theta}}
\newcommand{\E}{\mathbf{E}}
\newcommand{\X}{\mathbf{X}}
\newcommand{\Llambda}{\boldsymbol{\Lambda}}

The classical continuous-time Markov chain (CTMC) paradigm \cite{norris1998markov,anderson2012continuous} provides a rigorous formalism for random processes defined over discrete sets of abstract states and finds application throughout the mathematical sciences.  Chemists apply CTMC to the analysis of chemical reaction networks \cite{anderson2011continuous,anderson2012multilevel}, and computational biologists use them to model the evolution of genetic sequences \cite{suchard2001bayesian,ferreira2008bayesian,hobolth2009simulation}.  In computer science, CTMC provide the mathematical basis for queuing theory \cite{kendall1951some,kendall1953stochastic,yechiali1973queuing,cox2020queues,neuts2021structured}. 

Consider a state space $\mathcal{S}:=\{s_1,\dots,s_D\}$, where $D:=|\mathcal{S}|$ is the number of discrete items $s_d$. A CTMC is a stochastic process $\{x_t \in \mathcal{S}: t\geq 0\}$ that uses a $D\times D$ infinitesimal generator (rate) matrix $\QQ$ to parameterize each marginal probability $\ppi_t := (\Pr(x_t=s_1),\dots, \Pr(x_t=s_D))$ as $\ppi_t = \ppi_0 e^{t\QQ}$.
Here, each $\ppi_t$ is a row vector obtained by right multiplying an initial probability row vector $\ppi_0$ by the matrix exponential 
\begin{align}
    e^{t\QQ} := \sum_{i=0}^\infty \frac{(t \QQ)^i}{i!} \, .
\end{align}
Recent efforts model the rate matrix $\QQ$ itself as $\QQ = \QQ(\ttheta)$, where the dimensionality $K$ of the parameter vector $\ttheta\in \mathbb{R}^K$ may itself grow $\mathcal{O}(D^2)$ if it includes random effects associated to each element of $\QQ$.
The mixed-effects CTMC model of \cite{magee2024random} and \cite{didier2024surprising} parameterizes the off-diagonal elements of the rate matrix $\QQ$ as linear combinations of corresponding elements from $P$ fixed-effects matrices $\X_p$ and element-wise random effects $\lambda_{dd'}$:
\begin{align}\label{eq:glm}
\log [\QQ]_{dd'} =  \lambda_{dd'} +  \sum_{p=1}^P \beta_p  \left[\X_p \right]_{dd'}  \, , \quad d\neq d' \, , \quad  d,d' \in \{1,\dots,D\}\, .
\end{align}
Here, the parameter vector $\ttheta \in \mathbb{R}^K$ includes all random effects $\lambda_{dd'}$ and fixed-effect coefficients $\beta_p$, i.e., $K=\mathcal{O}(D^2)$. \cite{magee2024random} and \cite{didier2024surprising} place sparsity inducing Bayesian bridge \cite{polson2014bayesian} priors on the individual random effects $\lambda_{dd'}$ in order to encourage most random-effects to concentrate around zero while allowing some to obtain large values (indicating deviations from the fixed-effects model).  This strategy proves necessary because the $\mathcal{O}(D^2)$ number of random effects $\lambda_{dd'}$ can quickly grow larger than the number of observations.  For example, \cite{didier2024surprising} develop a phylogenetic extension to the basic mixed-effects CTMC model using 1,892 random effects and apply this model to the analysis of less than 300 observed SARS-CoV-2 cases.

\begin{figure}[!t]
    \centering
    \includegraphics[width=0.8\linewidth]{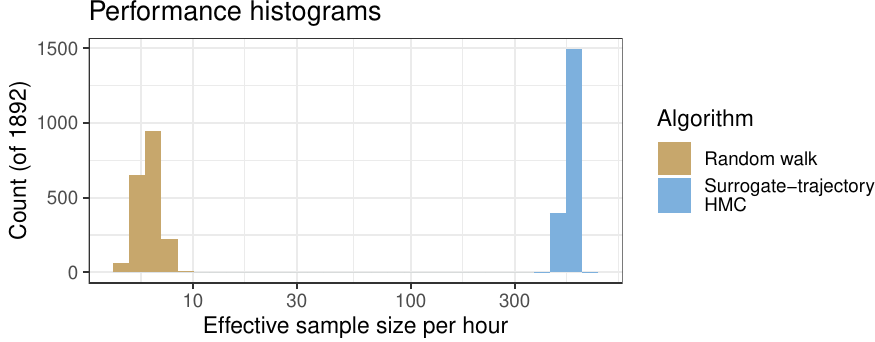}
    \caption{Inferring 1,892 random effects of a 44-dimensional continuous-time Markov chain (CTMC) model \eqref{eq:glm} within a larger phylogenetic CTMC model applied to the spread of SARS-CoV-2 between $D=44$ global regions.  The surrogate-trajectory HMC algorithm uses a first-order approximation to the matrix exponential derivative \eqref{eq:grad:series}. After searching over a field of leapfrog steps and target acceptance rates, we find that the relatively small number of 8 leapfrog steps and a target acceptance rate of 0.7 are optimal for this problem.  Due to the enormous $\mathcal{O}(D^7)$ time complexity of the exact gradient in this context, it is difficult to obtain effective sample sizes for classical HMC.}
    \label{fig:ctmcPerform}
\end{figure}

But inferring the posterior distribution of the high-dimensional parameter vector $\ttheta$ requires gradient-based MCMC approaches such as HMC.  Unfortunately, the gradient of the CTMC log-likelihood with respect to $\ttheta$ involves $D^2$ directional derivatives of the matrix exponential \cite{najfeld1995derivatives}
\begin{align}
    \nabla_{\E_{dd'}} e^{t\QQ} := \lim_{\epsilon \to 0} \frac{e^{t (\QQ+ \epsilon \E_{dd'})} - e^{t\QQ} }{\epsilon}  
 =  e^{t \QQ} \sum_{i = 0}^\infty \frac{t^{i+1}}{(i+1)!}  
	\left( \sum_{j = 0}^i (-1)^j \binom{i}{j} \QQ^j \E_{dd'} \QQ^{i -j}\right) \, . \label{eq:grad:series}
\end{align}
Here, the direction matrix $\E_{dd'}$ is the $D \times D$ matrix with $dd'$th entry equaling 1 and the rest equaling 0. Given the immense $\mathcal{O}(D^5)$ computational complexity of computing \eqref{eq:grad:series} for all $D^2$ directions $\E_{dd'}$, \cite{magee2024random} find it necessary to settle for the first-order approximation  $\widetilde{\nabla}_{\E_{dd'}}e^{t\QQ}:=te^{t\QQ}\E_{dd'}$ obtained by setting $i=0$ and find that the resulting approximate gradient with respect to $\ttheta$ serves surprisingly well within surrogate-trajectory HMC.  

Figure \ref{fig:ctmcPerform} compares performances of a random-walk algorithm and the surrogate HMC scheme of \cite{magee2024random} for modeling the spread of SARS-CoV-2 across 44 global regions. Within a random scan, the random-walk algorithm combines both multivariate adaptive random-walk Metropolis updates \cite{haario2001adaptive} and univariate random-walk updates. The two algorithms perform inference over the phylogenetic CTMC model's $44\times (44-1)=1892$ random effects, and we leave the other model parameters fixed at posterior modes for the purpose of comparison.  For surrogate-trajectory HMC, we tune the leapfrog step size using a target acceptance rate. After searching over a broad number of tuning parameter selections, we find that the relatively small number of 8 leapfrog steps and a target acceptance rate of 0.7 are optimal.  Finally, we note that using as few as 4 leapfrog steps also gives good results.  In this and the following section, we use the \textsc{CODA} \textsc{R} package \cite{coda,R} to calculate effective sample sizes for each dimension of each chain and \textsc{ggplot2} \cite{ggplot} to generate figures.

\subsection{Negative Example: Bayesian Multidimensional Scaling}\label{sec:bmds}

Multidimensional scaling (MDS) is a dimension reduction technique that maps pairwise dissimilarity measurements corresponding to a set of $N$ objects to a configuration of $N$ points within a low-dimensional Euclidean space \cite{torgerson1952cmds}. While classical MDS serves as a valuable data visualization tool, probabilistic extensions such as Bayesian multidimensional scaling (BMDS) \cite{ohraftery2001bmds} further enable uncertainty quantification in the context of Bayesian hierarchical models \cite{bedford2014, holbrook2021bigbmds, li2023}. BMDS models the set of $N$ objects' locations in low-dimensional space as latent variables conditioned on the observed dissimilarity measures which follow a prescribed joint probability distribution. Within BMDS, each observed dissimilarity measure $\delta_{nn'}$ equals the posited true measure $\delta_{nn'}^*$ plus a truncated Gaussian error:
\begin{align}
    \delta_{nn'} \sim \cN(\delta_{nn'}^*, \sigma^2) I(\delta_{nn'} > 0), \ n \ne n', \ n, n' \in \{1,...,N\} =: [N],
\end{align}
where $\delta^*_{nn'} = \sqrt{\sum_{d = 1}^D (x_{nd} - x_{n'd})^2}$ is the Euclidean distance between latent locations $\mathbf{x}_n, \,\mathbf{x}_{n'}$$\in\RR^D$, and $ I(\delta_{nn'} > 0)$ is the indicator function for $\delta_{nn'} > 0$. These assumptions yield the log-likelihood function 
\begin{align}
    \ell(\boldsymbol{\Delta}, \sigma^2) \propto -\frac{m}{2} \log(\sigma^2) - \sum_{n > n'} \biggr[ \frac{(\delta_{nn'} - \delta_{nn'}^*)^2}{2\sigma^2} + \log \Phi \biggr( \frac{\delta_{nn'}^*}{\sigma} \biggr) \biggr].
\end{align}    
Here, $\boldsymbol{\Delta}$ is the $N \times N$ matrix of observed dissimilarities, $m = N(N-1)/2$ is the number of dissimilarities and $\Phi(\cdot)$ is the standard normal cdf. For a Bayesian analysis of the model, one should specify priors for the unknown parameters, $\mathbf{X}$, the $N \times D$ matrix of unknown object locations and $\sigma^2$, the BMDS variance.

Using HMC to sample from the posterior distribution of $\X$ given $\boldsymbol{\Delta}$ requires calculating the gradient with respect to the latent locations. The log-likelihood gradient with respect to a single row $\x_n$ is 
\begin{align}
    \nabla{\x_n}  \ell(\boldsymbol{\Delta}, \sigma^2) = -\sum_{n \ne n'} \biggr[ \biggr( \frac{(\delta_{nn'}^* - \delta_{nn'})}{\sigma^2} + \frac{\phi(\delta_{nn'}^*/\sigma)}{\sigma \Phi(\delta_{nn'}^*/\sigma)} \biggr) \ \frac{(\mathbf{x}_n - \mathbf{x}_{n'})}{\delta_{nn'}^*} \biggr] \equiv -\sum_{n \ne n'} \mathbf{r}_{nn'}, \label{eq:grad:HMC}
\end{align}
where $\phi(\cdot)$ is the probability density function of a standard normal variate and $\mathbf{r}_{nn'}$ is the contribution of the $n'$th location to the gradient with respect to the $n$th location. The BMDS gradient (1.3.6) involves computing and summing over $\binom{N}{2}$ terms and requires $\mathcal{O}(N^2)$ floating point operations. Given the large number of calculations needed, the gradient becomes computationally cumbersome as the number of observations grows large. Here, we try using a small subset of the $\mathbf{r}_{nn'}$ terms within surrogate-trajectory HMC with a gradient approximation proposed by \cite{sheth2024sparse}. Instead of computing $(N - 1)$ dissimilarities per object, the sparse gradient includes $J_n \subset [N] \backslash \{n\}$ couplings per object $n$ where $J_n \ll N - 1$. One possibility for selecting object couplings is to extract $B$ bands or off-diagonals from the main diagonal of the observed distance matrix. The number of couplings $C$ is then the total number of elements in $B$ bands, i.e., $C = \sum_{b = 1}^B (N - b)$. Using this subset reduces the burden of computing the BMDS gradient to $\mathcal{O}(NB)$. 

\begin{figure}[!t]
    \centering
    \includegraphics[width=0.8\linewidth]{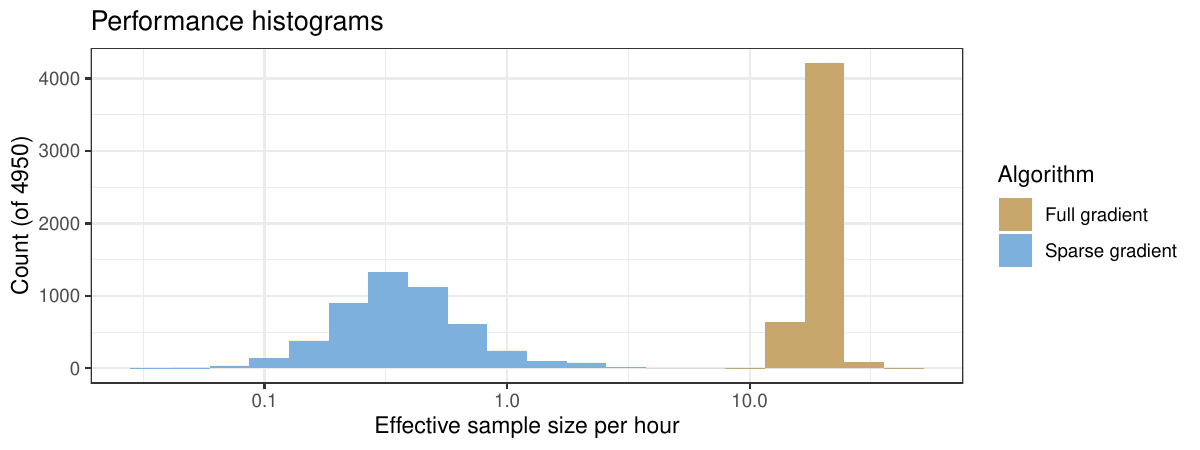}
    \caption{Effective sample sizes per hour of distances between inferred latent locations of 1,370 observed H1N1 influenza viruses. While we perform inference over all 1,370 viral locations, we only show performance results for a representative subset of 4,950 distances. Full HMC requires the entire BMDS gradient (\ref{eq:grad:HMC}), but surrogate-trajectory HMC uses a subset of the gradient selected by 50 bands. The cost-savings of surrogate-trajectory HMC fail to make up for its decreased sampling performance.}
    \label{fig:fluHMC}
\end{figure}

\cite{holbrook2021bigbmds} apply the BMDS framework to the phylogeographic analysis of the spread of H1N1 influenza using 1,370 viral samples. To scale BMDS to data of this size, they implement core model likelihood and log-likelihood gradient calculations on large graphics processing units and multi-core central processing units. Unfortunately, such an approach requires time-intensive coding and access to expensive computational hardware. We employ a similar analysis of the data, focusing on the performance of surrogate-trajectory HMC relative to that of generic HMC. 
In an HMC-within-Gibbs procedure, we use both algorithms to sample the latent location matrix $\mathbf{X}$, incorporating adaptive leapfrog step sizes to ensure the target acceptance rate is approximately 65\% while keeping the number of leapfrog steps fixed. We combine this approach with adaptive random-walk Metropolis-Hastings updates on the BMDS precision parameter, $1 / \sigma^2$ and Gibbs updates on the covariance structure that parameterizes H1N1's dispersal rate in the latent space \cite{holbrook2021bigbmds,sheth2024sparse}. Searching over tuning parameters to optimize surrogate-trajectory HMC's performance relative to that of HMC, we consider various numbers of bands for approximate gradient calculations $(B = 50, 100, 200, 700)$ with different numbers of leapfrog steps $(L = 4, 8, 20)$ and latent dimensions $(D = 2, 4)$. We observe marginal differences in performance between surrogate-trajectory HMC implementations with bands equaling 50, 100, 200 and 700: cost-savings generally fail to make up for losses in effective sample size.
We see the best performance of surrogate-trajectory HMC when we set the latent dimension to 2, the leapfrog steps to 8 and the number of bands to 50. Figure \ref{fig:fluHMC} compares the performance of surrogate-trajectory and generic HMC (also using 8 leapfrog steps) across 4,950 distances obtained by randomly selecting 100 locations from the set of 1,370 inferred latent locations. 

Despite reasonable attempts to tune surrogate-trajectory HMC, we do not find a case in which it outperforms traditional HMC. While not guaranteeing that a similar surrogate-trajectory approach will not beat HMC, this negative result highlights the difficulties that practitioners might face when designing their own surrogate-trajectory algorithms.  Finally, we note that \cite{sheth2024sparse} explore the combination of gradient approximations with corresponding likelihood approximations within the HMC accept-reject step. Although the resulting kernel does not leave its target invariant, \cite{sheth2024sparse} derive posterior consistency results in a simplified setting and demonstrate promising empirical performance.

\subsection{Bayesian Inversion of Nonlinear Partial Differential Equations}\label{sec:BIP}

Within Bayesian statistical inversion \cite{kaipio2005statistical,tarantola2005inverse}, one seeks to infer a parameter $\unk \in X$ through a forward model $\fwd$ from noisy observations $\data$ of its output: 
\begin{equation}
    \data = \fwd(\unk) + \ns,
\end{equation}
where $\ns$ is additive observational noise. Given a suitable prior
probability measure on the unknown parameter $\mpr(d\unk)$ and a
probability density $\mns$, the posterior measure is given by
\begin{equation}
    \mps(d\unk) = \frac{1}{Z}\mns(\data-\fwd(\unk))\mpr(d\unk), \text{ where } Z = \int_{X} \mns(\data-\fwd(\unk))\mpr(d\unk).
\end{equation}
In recent years, this statistical inversion methodology has been
developed in the context of PDE inverse problems where $\mathcal{G}$
involves solutions corresponding to an unknown infinite-dimensional
parameter as outlined in \cite{stuart2010inverse} and \cite{dashti2017bayesian}.  Thus, a key application of these statistical
inversion methods is in the estimation of unknown functions that serve
as physically meaningful unknown quantities in partial differential
equations, such as the space-varying diffusion constant in, e.g., a
model of groundwater flow; or a background fluid flow field
influencing the concentration of a passive solute
\cite{borggaard2020bayesian}; or in a fluid domain shape estimation
problem \cite{borggaard2023statistical}.

Of course these developments in the Bayesian approach to PDE inversion
have been made feasible only by the parallel development of the
Hilbert space sampling paradigm which we outline above in
\cref{sec:Hilbert}. Indeed, the pCN and $\infty$HMC algorithm have been
successfully employed to resolve complex, multimodal targets arising
from such PDE inverse problems in, e.g.,  \cite{borggaard2020bayesian,
  borggaard2023statistical}.  Here, on a per sample basis, $\infty$HMC
typically offers an order of magnitude better performance but these
seemingly dramatic advantages are typically hindered by the computational
cost and complexity of computing gradients.  While adjoint methods
provide a common method for resolving gradients of PDE-based forward
maps (see, e.g., \cite{hinze2009optimization} for an general overview
of the approach as applied to PDE-constrained optimization), in
practice, due to difficulties in derivation and other implementation
challenges, such methods may be unavailable.  Thus,
surrogate-trajectory methods (\cref{alg:inf:surgate}) and
multiproposal methods (\cref{alg:mpCN}) offer significant potential
to improve MCMC methods to the estimation of PDE
parameters.


As a paradigmatic test problem we consider here the estimatation of 
the flow of a fluid from measurements of a passive scalar (e.g., a dye
or solute) moving in the fluid. The physical model for this problem is
the advection-diffusion equation:
\begin{align}
  \partial_t \scf(t,\x) + \vf(\x) \cdot \nabla \scf(t,\x)
  = \diff \Delta \scf(t,\x), \quad \nabla \cdot \vf(\x) = 0
  \label{eq:ad:eqn}
\end{align}
on the periodic domain $\mathbb{T}^2=[0,1]^2$ where $\scf$ represents
the concentration of the solute, $\vf$ represents the unknown fluid
flow, and the initial condition $\scf_0(\x)=\scf(0,\x)$ and the
diffusion coefficient $\diff$ are assumed to be known. The goal, then,
is to estimate $\vf$---parameterized via its components in a
divergence-free Fourier representation---given noisy point
observational data, i.e.,
$\fwd(\vf)=\left( \scf(t_1,\x_1;\vf), \dots, \scf(t_N,\x_N;\vf)\right)$ and $\ns \sim \cN(0,\sigma^2 I)$.

\begin{figure}[!t]
    \centering
    \includegraphics[width=0.32\linewidth]{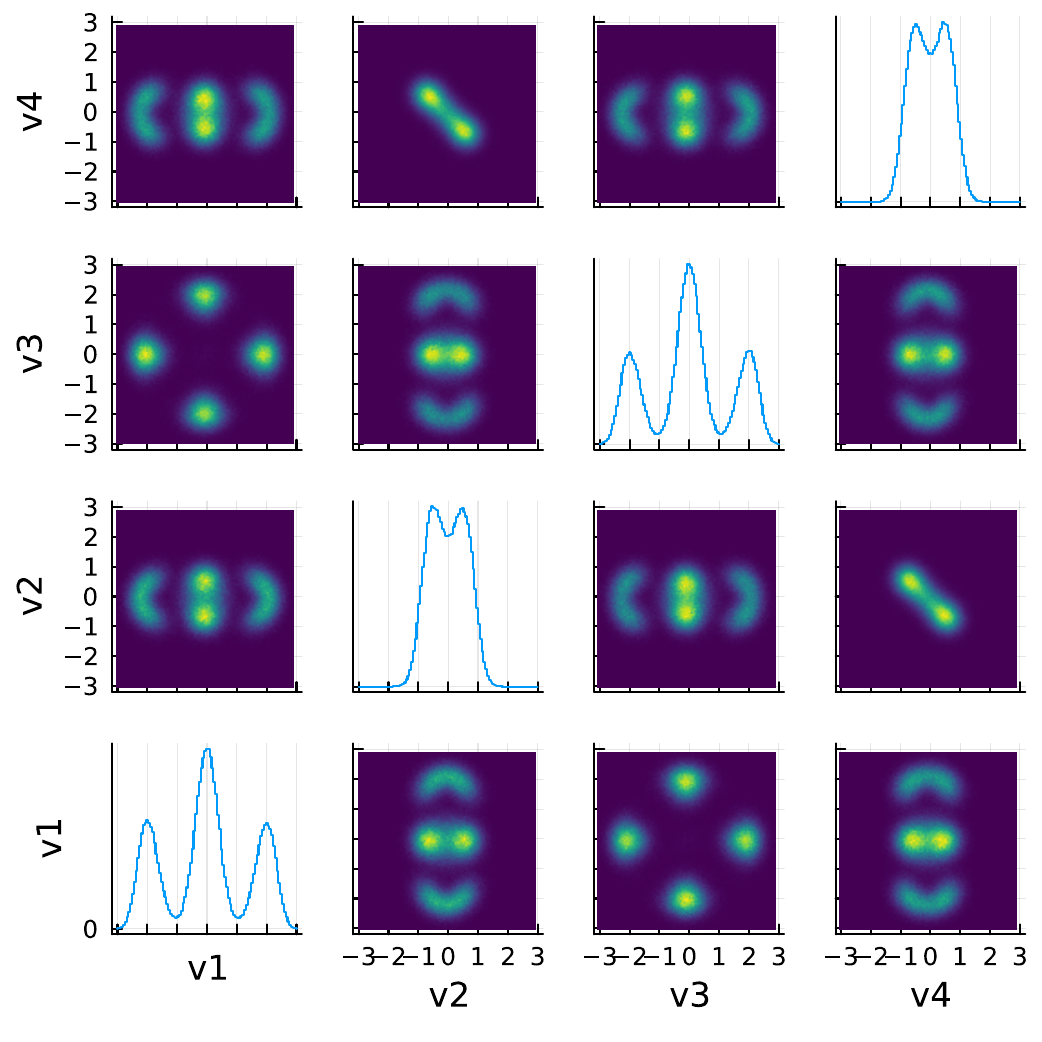}
    \hfill
    \includegraphics[width=0.32\linewidth]{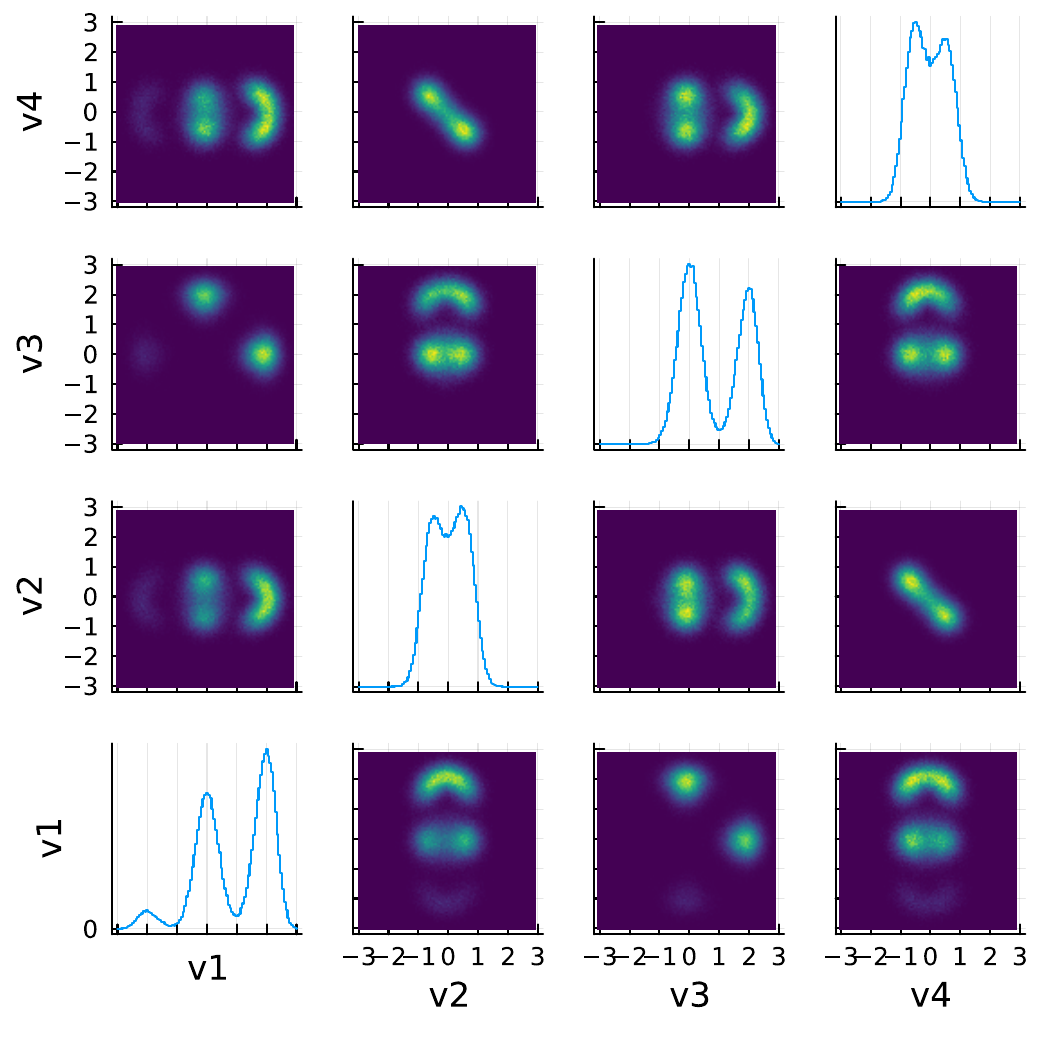}
    \hfill
    \includegraphics[width=0.32\linewidth]{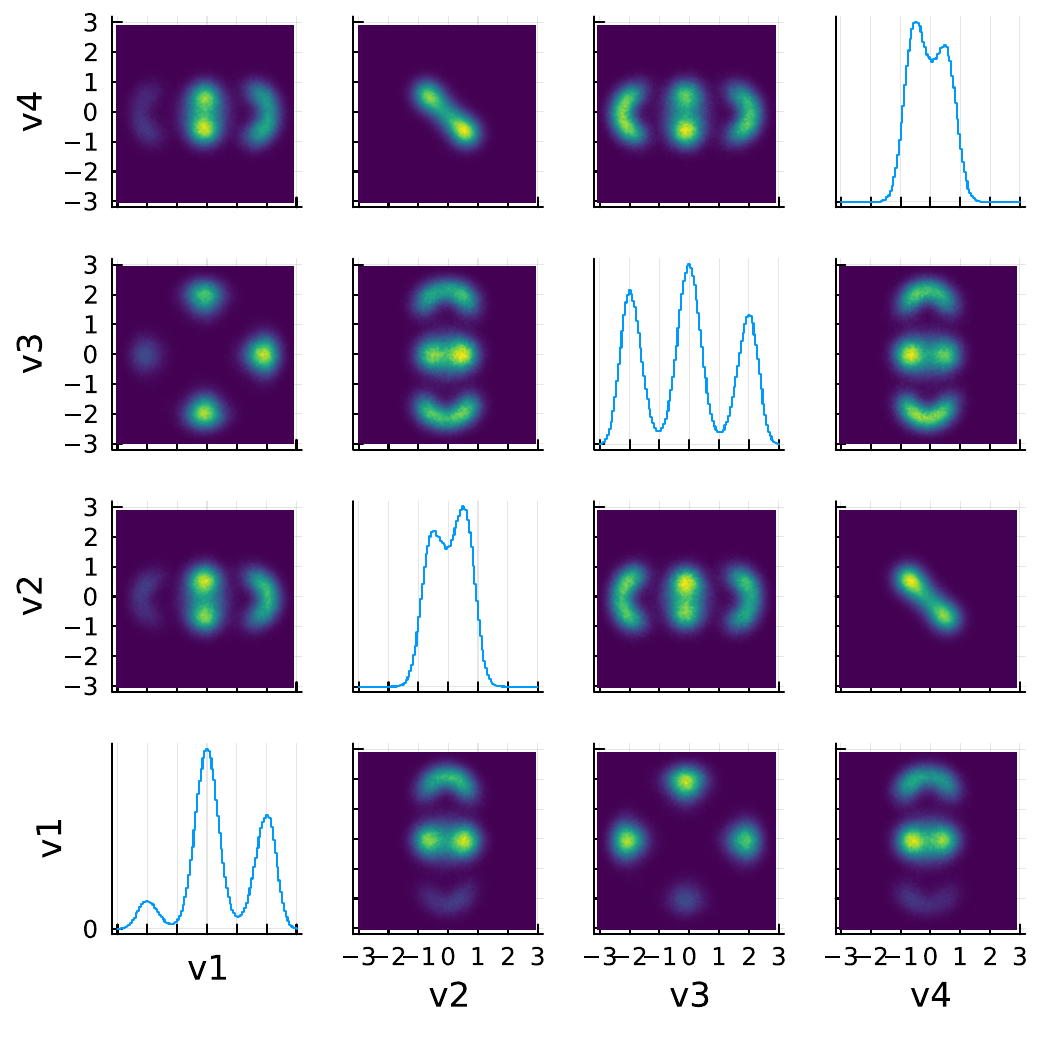}
        \vspace{-1em}
    \caption{Two-dimensional posterior density plots of the first four Fourier components of fluid flow $\vf$ for the advection-diffusion example \eqref{eq:ad:eqn}. Left: The ``true'' posterior from \cite{borggaard2020bayesian}. Middle: HMC after 200,000 samples only visits a subset of the posterior modes. Right: NNgHMC with the small neural network visits all modes after the computational equivalent of 200,000 HMC samples.}\label{fig:ad:2dhist}
\end{figure}

This problem features in \cite{krometis2018bayesian}, \cite{borggaard2019gpuaccelerated}, \cite{borggaard2020bayesian}, \cite{borggaard2020bayesiana} and \cite{glatt2024parallel}. Here, we consider the numerical example
introduced in \cite[Section 4.2]{borggaard2020bayesian}, which, due
to the symmetries of the problem, presents a challenging, multimodal
posterior to sample from (see the left plot of \cref{fig:ad:2dhist},
below). Because of the relatively simple boundary conditions, the PDE
solves required for the forward map can be computed efficiently,
enabling computation of a large number of MCMC samples in a reasonable
amount of time. The problem therefore represents a good case study of
algorithmic performance in the context of complex and high-dimensional
posterior structure.

One may apply gradient-based algorithms to this problem via an adjoint
method; as described in \cite[Section 3.3]{borggaard2020bayesian}. This requires solving a forced version of the forward map and then
computing a series of integrals. To illustrate the opportunity for
application of surrogate-trajectory methods to statistical inverse
problems governed by PDEs, here we use a neural network to approximate
the gradient as described in \cite{li2019neural}. As training data, we
take 10,000 samples from the approximate posterior computed via many
short HMC chains as described in \cite{borggaard2020bayesian}. We then
split them into 8,000 training samples and 2,000 samples used for
testing and train until the losses for the training and testing sets
began to diverge. We considered three neural network architectures:
\begin{enumerate}[label={(\arabic{enumi})}]
\item Small: three hidden layers of size 512, 1024, and 512 with
  softplus, relu, and softplus activation functions, respectively;
\item Medium: Five hidden layers of size 512 (softplus), 1024 (relu),
  2048 (softplus), 1024 (selu), and 512 (softplus);
\item Large: Seven hidden layers of size 512 (softplus), 1024 (relu),
  2048 (softplus), 4096 (selu), 2048 (softplus), 1024 (selu), and 512
  (softplus).
\end{enumerate}
We compare surrogate-trajectory HMC and MALA---denoted NNgHMC and
NNgMALA for neural network gradient HMC and MALA, respectively,
following the notation from \cite{li2019neural}---for these three
networks with (traditional) HMC and pCN. Finally, we compare also with
mpCN (\cref{sec:invol}) as an alternative gradient-free method.

\begin{figure}[!t]
    \centering
    \includegraphics[width=0.495\linewidth]{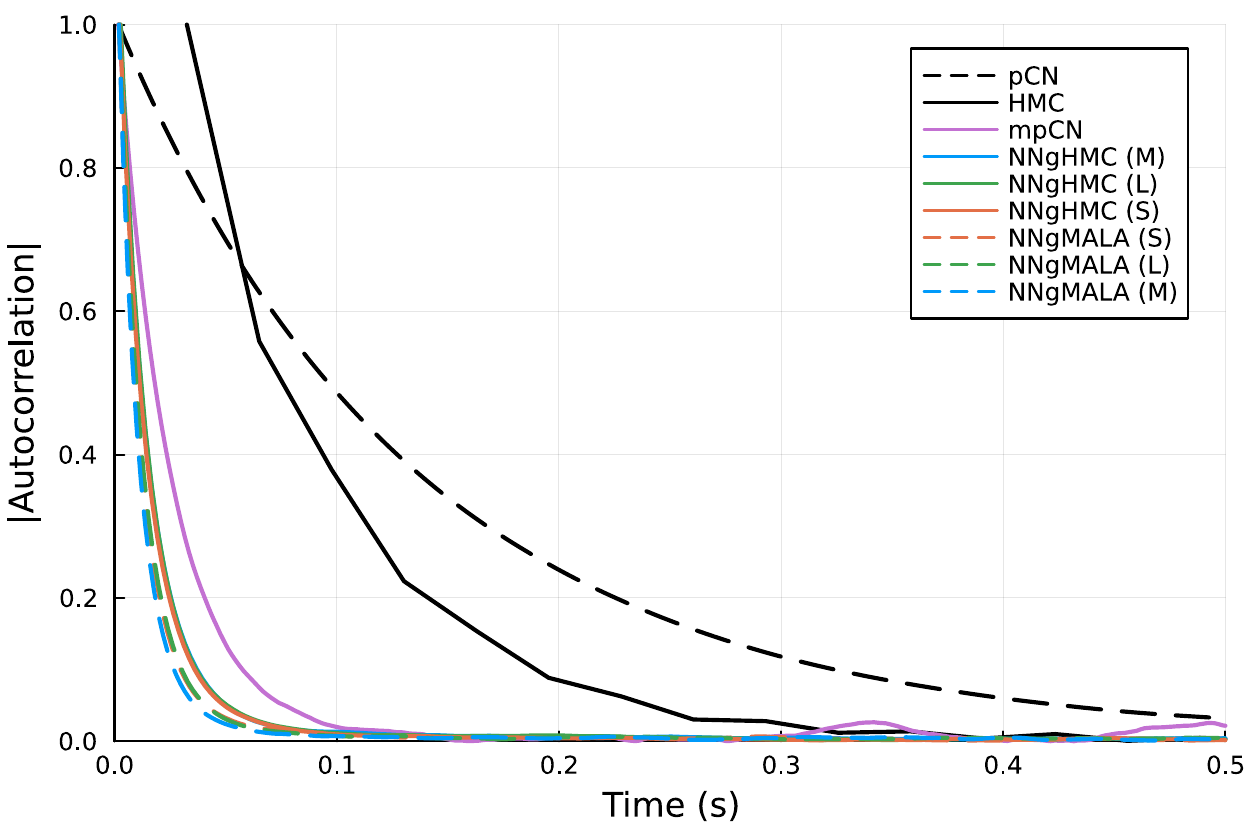}
    \hfill
    \includegraphics[width=0.495\linewidth]{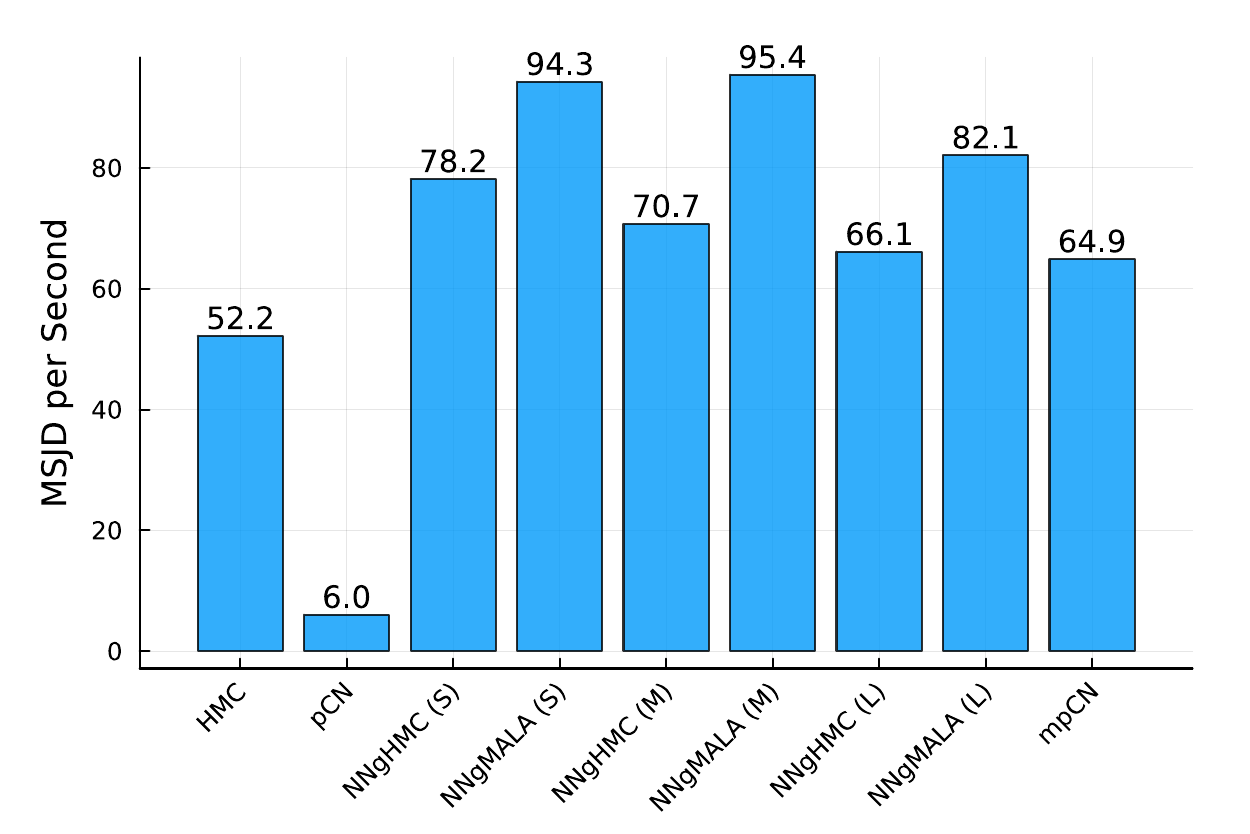}
    \vspace{-1.5em}
    \caption{Key performance results for the advection-diffusion example. All algorithms use tuning parameters that maximize mean squared jumping distance per second (MSJD/s). The left plot shows the absolute value of autocorrelation per second computed via the \textsc{StatsBase} \textsc{Julia} package \cite{bezanson2017julia}. The right plot shows mean squared jumping distance per second.  The mpCN results are synthetic insofar as they assume efficient parallelizaiton across target evaluations at all 64 proposals.}\label{fig:ad:ac_msjd}
\end{figure}

We generate chains for each method on a server with an AMD EPYC 7742
CPU and an A100 GPU. We perform neural network inference on the GPU,
reserving all other computations for the CPU, as we do not have a
GPU-accelerated advection-diffusion solver at hand. One benefit of
leveraging standard data-driven modeling techniques is the ready
availability of optimized libraries that can leverage high-performance
hardware. We choose the parameters selected for each method---step
size, number of steps for HMC, number of proposals for mpCN---by
running a parameter sweep of short chains and selecting values that
produce maximum mean-squared jumping distance (MSJD) per unit of
computational time.  \cref{fig:ad:ac_msjd} shows that the neural
network-based methods consistently produce lower autocorrelations and
larger mean jumps than either HMC or pCN when normalized by
computational costs; the performance mostly decays as the networks get
larger, as the increase in the computational costs outweighs the
improvement in the gradient approximation.  The neural network-based
methods are also generally competitive with HMC in the efficacy of
traversing the multimodal structure in Fourier modes 1 and 3:
\cref{fig:ad:2dhist} presents two-dimensional density plots for HMC
and NNgHMC (medium) with the ``true'' posterior distribution on the
first four Fourier modes. Finally, mpCN (with 64 proposals) appears to
provide a promising alternative when a gradient is not available; we refer the reader to \cite[Section 5.2.3]{glatt2024parallel} for tests of mpCN on an additional PDE-based inverse problem where a gradient is not available.

\section{Open Questions}\label{sec:open}

The surrogate-trajectory HMC literature is small but complicated.  On the one hand, the success of surrogate model-based approaches depends on the training and engineering prowess of the practitioner as well as the geometric idiosyncracies of the target. On the other hand, methods that truncate an analytic expression of the gradient intrinsically depend on the target itself, making generalization to other targets difficult.  Within both surrogate model and analytic truncation approaches, a large number of tuning parameter configurations present themselves.  At this time, surrogate-trajectory HMC methods stand as a collection of helpful hacks that have shown themselves to be useful in their own anecdotal scientific applications.

While we have made some progress on the theoretical foundations of surrogate-trajectory HMC, we believe that the growth of these methods beyond this unruly stage in their development will require the resolution of certain key questions. 
\begin{enumerate}
    \item Are there certain canonical targets for which one can obtain optimal scaling results?
\end{enumerate}
Such results would need to account for the relationships between the target distribution, its log-density gradient and a particular approximation of this gradient. These results would also need to account for the computational complexities associated with both the exact gradient and its approximation. Finally, these results would provide practitioners with an optimal combination of leapfrog step count and leapfrog step size for, e.g., the Gaussian mean of $N$ $D$-dimensional Gaussian observations with known covariance when the approximate gradient only makes use of this covariance's diagonal. Evidently, there is a tension between the specificity required to obtain quantitative results and the ability to generalize to other targets.
\begin{enumerate}
\setcounter{enumi}{1}
    \item Do certain instances of surrogate-trajectory HMC preserve properties, e.g., geometric ergodicity, obtained by HMC for certain targets?
\end{enumerate}
Again, it might be useful to begin with a simple target, for which we know HMC obtains geometric ergodicity \cite{livingstone2019geometric}.  Since this question does not involve computational complexity, one is free to consider a sequence of vector functions that converge to the log-posterior gradient \cite{li2019neural}. The goal would consist of establishing geometric ergodicity for \cref{alg:inf:surgate}, which in turn would require determining appropriate conditions on the approximation $\agp$ of the preconditioned log-posterior gradient. Here, one would hope to obtain a generalization of previous geometric ergodicity results derived for infinite-dimensional MCMC algorithms, such as pCN \cite{hairer2014spectral,durmus2016subgeometric}, $\infty$HMC \cite{pidstrigach2022convergence}, and versions of the $\infty$HMC algorithm in which the Hamiltonian dynamic integrates exactly \cite{glatt2021mixing,bou2021two}.


Bayesian inversion of PDEs is a rich application domain for \cref{alg:inf:surgate}
and \cref{alg:mpCN}.  The empirical studies of \cref{sec:BIP} and
\cite{glatt2024parallel} are preliminary, and more in-depth studies may point to greater performance advantages.  Within \cref{alg:inf:surgate},
the choice of a surrogate is an art, and we might ask in the `analytical'
case:
\begin{enumerate}
\setcounter{enumi}{2}
\item What are the most effective analytical approximations that
  should be used in \cref{alg:inf:surgate} for different PDE inverse
  problems?
\end{enumerate}
Addressing this question requires a model-by-model
approach, and the problems developed in
\cite{stuart2010inverse}, \cite{glatt2024parallel},
  \cite{borggaard2020bayesian} and \cite{borggaard2023statistical} provide a rich
testing ground.  For the problem of estimating velocity fields from solute
concentrations (\cref{sec:BIP}), one may approximate the adjoint method of \cite{borggaard2020bayesian} with a lower-order Galerkin truncation.  Other, more
artful, lower-order numerical methods may also prove useful here.

Stepping away from surrogate methods, we finish by asking one of the
possible questions that arise with multiproposal MCMC methods such as
Algorithm \ref{alg:mpCN}.    One avenue is to consider:
\begin{enumerate}
\setcounter{enumi}{3}
\item What is the effective kernel when one takes $p \to \infty$ in
  \eqref{eq:main:master:ker}  for different classes of
   multiproposal MCMC algorithms?
\end{enumerate}
Current work in progress suggests that this limit reveals further
interesting relationships between multiproposal and classical methods while
highlighting the salience of certain biased methods which do not
satisfy \eqref{eq:ar:RND:form}.  Behavior of $p=\infty$
kernels can also serve as an important object of study to determine
 theoretical limits of parallelization for corresponding finite-$p$ kernels.  Why expend computational resources if, for example, the multiproposal sampler collapses to a single-proposal sampler in the large-$p$ limit?

It is also interesting to consider situations in which one simultaneously takes both
the number of proposals $p$ and the number of dimensions $d$ to $\infty$. Such analyses may provide practical insights.
\begin{enumerate}
\setcounter{enumi}{4}
    \item Can one obtain optimal tuning results for certain multiproposal MCMC algorithms?
\end{enumerate}
Compared with the classical results surrounding optimal scaling for random-walk Metropolis, this question involves an additional tuning parameter, the number of proposals $p$. In practice, larger proposal counts appear necessary both for higher-dimensional problems and for more aggressive proposal distributions favoring proposals that are farther away from a current state. One might derive optimal scaling results in the large $p$ limit for both finite- and infinite-dimensional targets.  Finally, we note that multimodal distributions comprise a chief application for multiproposal methods and that these challenging targets typically require more---and more aggressive---proposals than do their unimodal counterparts.

\section*{Acknowledgments}
AJH is supported by grants NIH K25 AI153816, NSF DMS 2152774 and NSF
DMS 2236854. NEGH received support from NSF DMS 2108790. JAK received
support from NSF DMS-2108791. CFM is supported by NSF DMS 2009859 and
NSF DMS 2239325.

\bibliographystyle{alpha}
\bibliography{handbook}

\newcommand{\etalchar}[1]{$^{#1}$}
\begin{thebibliography}{GHHKM24}

\bibitem[AH12]{anderson2012multilevel}
David~F Anderson and Desmond~J Higham.
\newblock Multilevel monte carlo for continuous time markov chains, with
  applications in biochemical kinetics.
\newblock {\em Multiscale Modeling \& Simulation}, 10(1):146--179, 2012.

\bibitem[AK11]{anderson2011continuous}
David~F Anderson and Thomas~G Kurtz.
\newblock Continuous time markov chain models for chemical reaction networks.
\newblock In {\em Design and analysis of biomolecular circuits: engineering
  approaches to systems and synthetic biology}, pages 3--42. Springer, 2011.

\bibitem[ALL20]{andrieu2020general}
Christophe Andrieu, Anthony Lee, and Sam Livingstone.
\newblock A general perspective on the metropolis-hastings kernel.
\newblock {\em arXiv preprint arXiv:2012.14881}, 2020.

\bibitem[And12]{anderson2012continuous}
William~J Anderson.
\newblock {\em Continuous-time Markov chains: An applications-oriented
  approach}.
\newblock Springer Science \& Business Media, 2012.

\bibitem[Bar65]{barker1965monte}
Anthony~Alfred Barker.
\newblock Monte {C}arlo calculations of the radial distribution functions for a
  proton-electron plasma.
\newblock {\em Australian Journal of Physics}, 18(2):119--134, 1965.

\bibitem[BEKS17]{bezanson2017julia}
Jeff Bezanson, Alan Edelman, Stefan Karpinski, and Viral~B Shah.
\newblock Julia: A fresh approach to numerical computing.
\newblock {\em SIAM review}, 59(1):65--98, 2017.

\bibitem[BGHK19]{borggaard2019gpuaccelerated}
Jeff Borggaard, Nathan Glatt-Holtz, and Justin Krometis.
\newblock {GPU}-{Accelerated} {Particle} {Methods} for {Evaluation} of {Sparse}
  {Observations} for {Inverse} {Problems} {Constrained} by {Diffusion} {PDEs}.
\newblock {\em Journal of Computational Physics}, 2019.

\bibitem[BGHK20a]{borggaard2020bayesian}
Jeff Borggaard, Nathan Glatt-Holtz, and Justin Krometis.
\newblock A {Bayesian} {Approach} to {Estimating} {Background} {Flows} from a
  {Passive} {Scalar}.
\newblock {\em SIAM/ASA Journal on Uncertainty Quantification},
  8(3):1036--1060, January 2020.
\newblock Publisher: Society for Industrial and Applied Mathematics.

\bibitem[BGHK20b]{borggaard2020bayesiana}
Jeff Borggaard, Nathan Glatt-Holtz, and Justin Krometis.
\newblock On {Bayesian} {Consistency} for {Flows} {Observed} {Through} a
  {Passive} {Scalar}.
\newblock {\em Annals of Applied Probability}, 30(4):1762--1783, 2020.

\bibitem[BGHK23]{borggaard2023statistical}
Jeff Borggaard, Nathan~E Glatt-Holtz, and Justin Krometis.
\newblock A statistical framework for domain shape estimation in {S}tokes
  flows.
\newblock {\em Inverse Problems}, 2023.

\bibitem[BGL{\etalchar{+}}17]{beskos2017geometric}
Alexandros Beskos, Mark Girolami, Shiwei Lan, Patrick~E Farrell, and Andrew~M
  Stuart.
\newblock Geometric {MCMC} for infinite-dimensional inverse problems.
\newblock {\em Journal of Computational Physics}, 335:327--351, 2017.
\newblock Publisher: Elsevier.

\bibitem[BPSSS11]{beskos2011hybrid}
Alexandros Beskos, Frank~J Pinski, Jesús~Marıa Sanz-Serna, and Andrew~M
  Stuart.
\newblock Hybrid {Monte} {Carlo} on {Hilbert} spaces.
\newblock {\em Stochastic Processes and their Applications},
  121(10):2201--2230, 2011.
\newblock Publisher: Elsevier.

\bibitem[BRE21]{bou2021two}
Nawaf Bou-Rabee and Andreas Eberle.
\newblock {Two-scale coupling for preconditioned Hamiltonian Monte Carlo in
  infinite dimensions}.
\newblock {\em Stochastics and Partial Differential Equations: Analysis and
  Computations}, 9:207--242, 2021.

\bibitem[BRSS18]{bou2018geometric}
Nawaf Bou-Rabee and Jes{\'u}s~Maria Sanz-Serna.
\newblock Geometric integrators and the {Hamiltonian Monte Carlo} method.
\newblock {\em Acta Numerica}, 27:113--206, 2018.

\bibitem[BRSV08]{beskos2008mcmc}
A.~Beskos, G.~Roberts, A.~Stuart, and J.~Voss.
\newblock {MCMC methods for diffusion bridges}.
\newblock {\em Stochastics and Dynamics}, 8(03):319--350, 2008.

\bibitem[BSL{\etalchar{+}}14]{bedford2014}
Trevor Bedford, Marc~A. Suchard, Philippe Lemey, Gytis Dudas, Victoria Gregory,
  Alan~J. Hay, John~W. McCauley, Colin~A. Russel, Derek~J. Smith, and Andrew
  Rambaut.
\newblock Integrating influenza antigenic dynamics with molecular evolution.
\newblock {\em eLife}, 3:e01914, 2014.

\bibitem[Cal14]{calderhead2014general}
B.~Calderhead.
\newblock {A general construction for parallelizing Metropolis-Hastings
  algorithms}.
\newblock {\em Proceedings of the National Academy of Sciences},
  111(49):17408--17413, 2014.

\bibitem[Cox20]{cox2020queues}
DR~Cox.
\newblock {\em Queues}.
\newblock Chapman and Hall/CRC, 2020.

\bibitem[CRSW13]{cotter2013mcmc}
Simon~L Cotter, Gareth~O Roberts, Andrew~M Stuart, and David White.
\newblock {MCMC} methods for functions: modifying old algorithms to make them
  faster.
\newblock {\em Statistical Science}, 28(3):424--446, 2013.
\newblock Publisher: Institute of Mathematical Statistics.

\bibitem[DFM16]{durmus2016subgeometric}
Alain Durmus, Gersende Fort, and \'Eric Moulines.
\newblock {Subgeometric rates of convergence in Wasserstein distance for Markov
  chains}.
\newblock {\em Annales de l'Institut Henri Poincar{\'e}, Probabilit{\'e}s et
  Statistiques}, 52(4):1799--1822, 2016.

\bibitem[DGHH{\etalchar{+}}24]{didier2024surprising}
Gustavo Didier, Nathan~E Glatt-Holtz, Andrew~J Holbrook, Andrew~F Magee, and
  Marc~A Suchard.
\newblock On the surprising effectiveness of a simple matrix exponential
  derivative approximation, with application to global sars-cov-2.
\newblock {\em Proceedings of the National Academy of Sciences},
  121(3):e2318989121, 2024.

\bibitem[DJ09]{delmas2009does}
J.-F. Delmas and B.~Jourdain.
\newblock {Does waste recycling really improve the multi-proposal
  Metropolis--Hastings algorithm? An analysis based on control variates}.
\newblock {\em Journal of applied probability}, 46(4):938--959, 2009.

\bibitem[DKPR87]{duane1987hybrid}
Simon Duane, Anthony~D Kennedy, Brian~J Pendleton, and Duncan Roweth.
\newblock Hybrid {Monte Carlo}.
\newblock {\em Physics letters B}, 195(2):216--222, 1987.

\bibitem[DPZ14]{da2014stochastic}
Giuseppe Da~Prato and Jerzy Zabczyk.
\newblock {\em Stochastic equations in infinite dimensions}.
\newblock Cambridge university press, 2014.

\bibitem[DS17]{dashti2017bayesian}
Masoumeh Dashti and Andrew~M Stuart.
\newblock The {Bayesian} approach to inverse problems.
\newblock {\em Handbook of Uncertainty Quantification}, pages 311--428, 2017.
\newblock Publisher: Springer.

\bibitem[Fre04]{frenkel2004speed}
D.~Frenkel.
\newblock {Speed-up of Monte Carlo simulations by sampling of rejected states}.
\newblock {\em Proceedings of the National Academy of Sciences},
  101(51):17571--17575, 2004.

\bibitem[FS08]{ferreira2008bayesian}
Marco~AR Ferreira and Marc~A Suchard.
\newblock Bayesian analysis of elapsed times in continuous-time {M}arkov
  chains.
\newblock {\em Canadian Journal of Statistics}, 36(3):355--368, 2008.

\bibitem[GC11]{girolami2011riemann}
Mark Girolami and Ben Calderhead.
\newblock Riemann manifold {Langevin and Hamiltonian Monte Carlo} methods.
\newblock {\em Journal of the Royal Statistical Society Series B: Statistical
  Methodology}, 73(2):123--214, 2011.

\bibitem[Gey11]{geyer2011introduction}
Charles~J Geyer.
\newblock Introduction to {Markov Chain Monte Carlo}.
\newblock In {\em Handbook of Markov Chain Monte Carlo}, pages 3--48. CRC
  Press, 2011.

\bibitem[GHHKM24]{glatt2024parallel}
Nathan~E Glatt-Holtz, Andrew~J Holbrook, Justin~A Krometis, and Cecilia~F
  Mondaini.
\newblock Parallel {MCMC} algorithms: Theoretical foundations, algorithm
  design, case studies.
\newblock {\em Transactions of Mathematics and Its Applications}, 8(2):tnae004,
  2024.

\bibitem[GHKM23]{glatt2023accept}
Nathan Glatt-Holtz, Justin Krometis, and Cecilia Mondaini.
\newblock On the accept--reject mechanism for {Metropolis--Hastings}
  algorithms.
\newblock {\em The Annals of Applied Probability}, 33(6B):5279--5333, 2023.

\bibitem[GHM21]{glatt2021mixing}
Nathan~E Glatt-Holtz and Cecilia~F Mondaini.
\newblock {Mixing rates for Hamiltonian Monte Carlo algorithms in finite and
  infinite dimensions}.
\newblock {\em Stochastics and Partial Differential Equations: Analysis and
  Computations}, pages 1--74, 2021.

\bibitem[Gre95]{green1995reversible}
Peter~J Green.
\newblock Reversible jump {Markov chain Monte Carlo computation and Bayesian
  model determination}.
\newblock {\em Biometrika}, 82(4):711--732, 1995.

\bibitem[Hai09]{hairer2009introduction}
Martin Hairer.
\newblock An introduction to stochastic pdes.
\newblock {\em arXiv preprint arXiv:0907.4178}, 2009.

\bibitem[HG{\etalchar{+}}14]{hoffman2014no}
Matthew~D Hoffman, Andrew Gelman, et~al.
\newblock The no-u-turn sampler: adaptively setting path lengths in hamiltonian
  monte carlo.
\newblock {\em J. Mach. Learn. Res.}, 15(1):1593--1623, 2014.

\bibitem[HLB{\etalchar{+}}21]{holbrook2021bigbmds}
Andrew~J. Holbrook, Philippe Lemey, Guy Baele, Simon Dellicour, Dirk Brockmann,
  Andrew Rambaut, and Marc~A. Suchard.
\newblock Massive parallelization boosts big {B}ayesian multidimensional
  scaling.
\newblock {\em Journal of Computational and Graphical Statistics},
  30(1):11--24, 2021.

\bibitem[Hol23a]{Holbrook2023generating}
Andrew~J Holbrook.
\newblock Generating mcmc proposals by randomly rotating the regular simplex.
\newblock {\em Journal of Multivariate Analysis}, 194:105106, 2023.

\bibitem[Hol23b]{holbrook2023quantum}
Andrew~J. Holbrook.
\newblock A quantum parallel markov chain monte carlo.
\newblock {\em Journal of Computational and Graphical Statistics},
  32(4):1402--1415, 2023.

\bibitem[HPUU09]{hinze2009optimization}
Michael Hinze, René Pinnau, Michael Ulbrich, and Stefan Ulbrich.
\newblock {\em Optimization with {PDE} {Constraints}}, volume~23 of {\em
  Mathematical {Modelling}: {Theory} and {Applications}}.
\newblock Springer, New York, 2009.

\bibitem[HS09]{hobolth2009simulation}
Asger Hobolth and Eric~A Stone.
\newblock Simulation from endpoint-conditioned, continuous-time markov chains
  on a finite state space, with applications to molecular evolution.
\newblock {\em The annals of applied statistics}, 3(3):1204, 2009.

\bibitem[HST01]{haario2001adaptive}
Heikki Haario, Eero Saksman, and Johanna Tamminen.
\newblock An adaptive {M}etropolis algorithm.
\newblock {\em Bernoulli}, pages 223--242, 2001.

\bibitem[HSV14]{hairer2014spectral}
Martin Hairer, Andrew~M Stuart, and Sebastian~J Vollmer.
\newblock Spectral gaps for a {M}etropolis--{H}astings algorithm in infinite
  dimensions.
\newblock {\em The Annals of Applied Probability}, 24(6):2455--2490, 2014.

\bibitem[Ken51]{kendall1951some}
David~G Kendall.
\newblock Some problems in the theory of queues.
\newblock {\em Journal of the Royal Statistical Society: Series B
  (Methodological)}, 13(2):151--173, 1951.

\bibitem[Ken53]{kendall1953stochastic}
David~G Kendall.
\newblock Stochastic processes occurring in the theory of queues and their
  analysis by the method of the imbedded markov chain.
\newblock {\em The Annals of Mathematical Statistics}, pages 338--354, 1953.

\bibitem[Kro18]{krometis2018bayesian}
Justin Krometis.
\newblock {\em A {Bayesian} {Approach} to {Estimating} {Background} {Flows}
  from a {Passive} {Scalar}}.
\newblock {PhD} {Thesis}, Virginia Polytechnic Institute and State University,
  2018.

\bibitem[KS05]{kaipio2005statistical}
Jari Kaipio and Erkki Somersalo.
\newblock {\em Statistical and {Computational} {Inverse} {Problems}}, volume
  160 of {\em Applied {Mathematical} {Sciences}}.
\newblock Springer Science \& Business Media, 2005.

\bibitem[LBBG19]{livingstone2019geometric}
Samuel Livingstone, Micharl Betancourt, Simon Byrne, and Mark Girolami.
\newblock On the geometric ergodicity of hamiltonian monte carlo.
\newblock {\em Bernoulli}, 25(4A):3109--3138, 2019.

\bibitem[LBTCG16]{lan2016emulation}
Shiwei Lan, Tan Bui-Thanh, Mike Christie, and Mark Girolami.
\newblock Emulation of higher-order tensors in manifold monte carlo methods for
  bayesian inverse problems.
\newblock {\em Journal of Computational Physics}, 308:81--101, 2016.

\bibitem[LGH{\etalchar{+}}23]{li2023}
YQ~Li, M~Ghafari, AJ~Holbrook, I~Boonen, N~Amor, S~Catalano, JP~Webster, YY~Li,
  HT~Li, V~Vergote, P~Maes, YL~Chong, A~Laudisoit, P~Baelo, S~Ngoy,
  SG~Mbalitini, GC~Gembu, AP~Musaba, J~Goüy~de Bellocq, H~Leirs, E~Verheyen,
  OG~Pybus, A~Katzourakis, AN~Alagaili, S~Gryseels, YC~Li, MA~Suchard,
  M~Bletsa, and P~Lemey.
\newblock The evolutionary history of hepaciviruses.
\newblock {\em bioRxiv : the preprint server for biology}, 2023.

\bibitem[LHSB19]{li2019neural}
Lingge Li, Andrew Holbrook, Babak Shahbaba, and Pierre Baldi.
\newblock Neural network gradient {Hamiltonian Monte Carlo}.
\newblock {\em Computational statistics}, 34:281--299, 2019.

\bibitem[LLW00]{liu2000multiple}
J.S. Liu, F.~Liang, and W.H. Wong.
\newblock {The multiple-try method and local optimization in Metropolis
  sampling}.
\newblock {\em Journal of the American Statistical Association},
  95(449):121--134, 2000.

\bibitem[LT19]{luo2019multiple}
X.~Luo and H.~Tjelmeland.
\newblock {A multiple-try Metropolis--Hastings algorithm with tailored
  proposals}.
\newblock {\em Computational Statistics}, 34(3):1109--1133, 2019.

\bibitem[MHP{\etalchar{+}}24]{magee2024random}
Andrew~F Magee, Andrew~J Holbrook, Jonathan~E Pekar, Itzue~W Caviedes-Solis,
  Fredrick~A Matsen~IV, Guy Baele, Joel~O Wertheim, Xiang Ji, Philippe Lemey,
  and Marc~A Suchard.
\newblock Random-effects substitution models for phylogenetics via scalable
  gradient approximations.
\newblock {\em Systematic Biology}, page syae019, 2024.

\bibitem[Nea99]{neal1999regression}
Radford~M. Neal.
\newblock Regression and classification using gaussian process priors (with
  discussion).
\newblock {\em Bayesian statistics 6}, pages 475--501, 1999.

\bibitem[Nea03]{neal2003markov}
R.M. Neal.
\newblock {Markov chain sampling for non-linear state space models using
  embedded hidden Markov models}.
\newblock {\em arXiv preprint math/0305039}, 2003.

\bibitem[Nea11]{neal2011mcmc}
Radford~M Neal.
\newblock {MCMC Using Hamiltonian Dynamics}.
\newblock In {\em Handbook of Markov Chain Monte Carlo}, pages 113--162. CRC
  Press, 2011.

\bibitem[Neu21]{neuts2021structured}
Marcel~F Neuts.
\newblock {\em Structured stochastic matrices of M/G/1 type and their
  applications}.
\newblock CRC Press, 2021.

\bibitem[NH95]{najfeld1995derivatives}
Igor Najfeld and Timothy~F Havel.
\newblock Derivatives of the matrix exponential and their computation.
\newblock {\em Advances in applied mathematics}, 16(3):321--375, 1995.

\bibitem[Nor98]{norris1998markov}
James~R Norris.
\newblock {\em Markov chains}.
\newblock Cambridge University Press, 1998.

\bibitem[NWEV20]{neklyudov2020involutive}
Kirill Neklyudov, Max Welling, Evgenii Egorov, and Dmitry Vetrov.
\newblock Involutive mcmc: a unifying framework.
\newblock In {\em International Conference on Machine Learning}, pages
  7273--7282. PMLR, 2020.

\bibitem[OR01]{ohraftery2001bmds}
Man-Suk Oh and Adrian~E. Raftery.
\newblock Bayesian multidimensional scaling and choice of dimension.
\newblock {\em Journal of the American Statistical Association},
  96(455):1031--1044, 2001.

\bibitem[PBCV06]{coda}
Martyn Plummer, Nicky Best, Kate Cowles, and Karen Vines.
\newblock Coda: Convergence diagnosis and output analysis for mcmc.
\newblock {\em R News}, 6(1):7--11, 2006.

\bibitem[Pes73]{peskun1973optimum}
Peter~H Peskun.
\newblock Optimum {Monte-Carlo sampling using Markov Chains}.
\newblock {\em Biometrika}, 60(3):607--612, 1973.

\bibitem[Pid22]{pidstrigach2022convergence}
Jakiw Pidstrigach.
\newblock {Convergence of preconditioned Hamiltonian Monte Carlo on Hilbert
  spaces}.
\newblock {\em IMA Journal of Numerical Analysis}, 43(5):2665--2713, 2022.

\bibitem[PSW14]{polson2014bayesian}
Nicholas~G Polson, James~G Scott, and Jesse Windle.
\newblock The bayesian bridge.
\newblock {\em Journal of the Royal Statistical Society: Series B: Statistical
  Methodology}, pages 713--733, 2014.

\bibitem[{R C}23]{R}
{R Core Team}.
\newblock {\em R: A Language and Environment for Statistical Computing}.
\newblock R Foundation for Statistical Computing, Vienna, Austria, 2023.

\bibitem[Ras03]{rasmussen2003gaussian}
Carl~Edward Rasmussen.
\newblock Gaussian processes to speed up hybrid {Monte Carlo for expensive
  Bayesian integrals}.
\newblock In {\em Seventh Valencia international meeting, dedicated to Dennis
  V. Lindley}, pages 651--659. Oxford University Press, 2003.

\bibitem[SC21]{schwedes2021rao}
T.~Schwedes and B.~Calderhead.
\newblock {Rao-Blackwellised parallel MCMC}.
\newblock In {\em International Conference on Artificial Intelligence and
  Statistics}, pages 3448--3456. PMLR, 2021.

\bibitem[SSH24]{sheth2024sparse}
Ami Sheth, Aaron Smith, and Andrew~J Holbrook.
\newblock Sparse bayesian multidimensional scaling (s).
\newblock {\em arXiv preprint arXiv:2406.15573}, 2024.

\bibitem[Stu10]{stuart2010inverse}
Andrew~M Stuart.
\newblock Inverse problems: a {Bayesian} perspective.
\newblock {\em Acta Numerica}, 19:451--559, 2010.
\newblock Publisher: Cambridge University Press.

\bibitem[SWS01]{suchard2001bayesian}
Marc~A Suchard, Robert~E Weiss, and Janet~S Sinsheimer.
\newblock Bayesian selection of continuous-time markov chain evolutionary
  models.
\newblock {\em Molecular biology and evolution}, 18(6):1001--1013, 2001.

\bibitem[Tar05]{tarantola2005inverse}
Albert Tarantola.
\newblock {\em Inverse {Problem} {Theory} and {Methods} for {Model} {Parameter}
  {Estimation}}.
\newblock SIAM, 2005.

\bibitem[Tie98]{Tierney1998}
L.~Tierney.
\newblock A note on {M}etropolis-{H}astings kernels for general state spaces.
\newblock {\em Annals of Applied Probability}, pages 1--9, 1998.

\bibitem[Tje04]{tjelmeland2004using}
H.~Tjelmeland.
\newblock {Using all Metropolis--Hastings proposals to estimate mean values}.
\newblock Technical report, Citeseer, 2004.

\bibitem[Tor52]{torgerson1952cmds}
Warren~S. Torgerson.
\newblock Multidimensional scaling: I. theory and method.
\newblock {\em Psychometrika}, 17(4):401--419, 1952.

\bibitem[Wic16]{ggplot}
Hadley Wickham.
\newblock {\em ggplot2: Elegant Graphics for Data Analysis}.
\newblock Springer-Verlag New York, 2016.

\bibitem[Yec73]{yechiali1973queuing}
Uri Yechiali.
\newblock A queuing-type birth-and-death process defined on a continuous-time
  markov chain.
\newblock {\em Operations Research}, 21(2):604--609, 1973.

\bibitem[ZSZ17a]{zhang2017hamiltonian}
Cheng Zhang, Babak Shahbaba, and Hongkai Zhao.
\newblock Hamiltonian monte carlo acceleration using surrogate functions with
  random bases.
\newblock {\em Statistics and computing}, 27:1473--1490, 2017.

\bibitem[ZSZ17b]{zhang2017precomputing}
Cheng Zhang, Babak Shahbaba, and Hongkai Zhao.
\newblock Precomputing strategy for hamiltonian monte carlo method based on
  regularity in parameter space.
\newblock {\em Computational Statistics}, 32:253--279, 2017.

\bibitem[ZSZ18]{zhang2018variational}
Cheng Zhang, Babak Shahbaba, and Hongkai Zhao.
\newblock Variational hamiltonian monte carlo via score matching.
\newblock {\em Bayesian analysis}, 13(2):485, 2018.

\end{thebibliography}

\end{document}